\documentclass[trackchanges, twocolumn]{aastex701} %linenumbers,

\usepackage{comment}
\begin{document}

%\title{Direct Optical Evidence of Late-Stage Infall in AB Aurigae: An Anomalous [O I] Inflow and a Crushed Magnetosphere}

%\title{Probing the Inner Boundary of Late-Stage Infall in AB Aurigae: A Crushed Magnetosphere and a Stagnant [O I] Reservoir}
\title{Direct Optical Evidence of Late-Stage Infall in AB Aurigae: A Stagnant [O~I] Reservoir and a Crushed Magnetosphere}

\correspondingauthor{Dipen Sahu}
\email{dsahu@prl.res.in}

\author[0000-0002-4393-3463]{Dipen Sahu}
\email{dsahu@prl.res.in}
\affiliation{Physical Research Laboratory, Navrangpura, Ahmedabad, Gujarat 380009, India}

\author[0000-0001-8983-5300]{Rishikesh Sharma}
\email{rishikesh@prl.res.in}
\affiliation{Physical Research Laboratory, Navrangpura, Ahmedabad, Gujarat 380009, India}

\author[0009-0006-9996-1814]{Shubhendra Nath Das}
\email{shubhendra@prl.res.in}
\affiliation{Physical Research Laboratory, Navrangpura, Ahmedabad, Gujarat 380009, India}
\affiliation{Physics Department, Indian Institute of Technology Gandhinagar, Gandhinagar, Gujarat 382355, India}

\author[0000-0002-3815-8407]{Abhijit Chakraborty}
\email{abhijit@prl.res.in}
\affiliation{Physical Research Laboratory, Navrangpura, Ahmedabad, Gujarat 380009, India}

\author[0000-0002-3913-3746]{Justyn Campbell-White}
\email{astrojustyn@gmail.com}
\affiliation{European Southern Observatory, Karl-Schwarzschild-Strasse 2, 85748 Garching bei M\"{u}nchen, Germany}

%\collaboration{all}{The PARAS team }

%% Use the \collaboration command to identify collaborations. This command
%% takes an optional argument that is either a number or the word "all"
%% which tells the compiler how many of the authors above the command to
%% show. For example "\collaboration[all]{(DELVE Collaboration)}" wil include
%% all the authors above this command.
%%
%% Mark off the abstract in the ``abstract'' environment. 
\begin{abstract}
%%%%%updated abstract
Massive planet-carved cavities in transition disks should theoretically throttle inward gas transport, challenging our understanding of how central stars maintain vigorous accretion. To investigate how macro-scale late-stage infall traverses these gaps, we present multi-epoch, extreme-resolution ($R \approx 107{,}000$) PARAS-2 optical spectroscopy of the benchmark Herbig Ae system AB Aurigae. By resolving the kinematics of H$\alpha$, He~I $\lambda5876$, [O~I] $\lambda6300, 6363$, and Na~I~D, we map the innermost accretion environment.  We find that the [O~I] emission is centered near the stellar rest velocity with symmetric broadening of $\sim 35$ km s$^{-1}$. Restricted to $T \lesssim 3800$ K, this profile traces a stagnant, gravitationally bound Keplerian gas reservoir at $\sim 1$ au. Therefore, it provides a strong optical evidence that late-stage infall accumulates in an inner gas reservoir and subsequently feeds the innermost dust cavity. From this reservoir, gas is transported inward and crashes onto the star, driving a highly active accretion rate of $\dot{M}_{\rm acc} \approx 4 \times 10^{-7}\,M_\odot\,{\rm yr}^{-1}$. The associated ram pressure crushes the stellar magnetosphere to $R_{\rm mag} \approx 1.2\,R_\star$, which explains the restricted He~I free-fall velocities and the highly variable inner wind. We also isolate a stable, slow H$\alpha$ wind component, likely tracing an extended photoevaporative disk wind.

\end{abstract}

%% Keywords should appear after the \end{abstract} command. 
%% The AAS Journals now uses Unified Astronomy Thesaurus (UAT) concepts:
%% https://astrothesaurus.org
%% You will be asked to selected these concepts during the submission process
%% but this old "keyword" functionality is maintained in case authors want
%% to include these concepts in their preprints.
%%
%% You can use the \uat command to link your UAT concepts back its source.

\keywords{\uat{Protoplanetary disks}{1300} --- \uat{Herbig Ae/Be stars}{723} --- \uat{Stellar accretion}{1578} --- \uat{Stellar accretion disks}{1579} --- \uat{Pre-main sequence stars}{1290} --- \uat{High resolution spectroscopy}{2096}}

\section{Introduction} \label{sec:intro} 
%=======
% 1. Set up the problem: Transition disks and massive cavities 

The evolution and dispersal of protoplanetary disks govern the timescale and environment of planet formation. A critical phase in this evolution is the transition disk stage, characterized by a massive outer dust disk and a significantly depleted inner dust cavity \citep[e.g.,][]{Alexander2014prpl.conf..475A}. Despite these massive inner clearings, which often extend to tens or hundreds of au, many transition disks continue to exhibit robust signatures of gas accretion onto the central star \citep{Mendigutia2012A&A...543A..59M, Manara2014A&A...568A..18M, Mentigutia2026arXiv260405040M}. Resolving exactly how gas dynamically traverses these massive planet-carved cavities to replenish the inner disk (au an sub-au scale) and maintain active accretion remains a central challenge in modern astrophysics. 

Recent high-contrast imaging and interferometry have revealed complex non-axisymmetric structures, such as cavity-crossing spiral arms, which are theorized to act as mass-transport bridges \citep{Tang2017ApJ...840...32T, Boccaletti2020A&A...637L...5B}. However, the ultimate kinematic impact of this material as it arrives at the innermost star-disk interface is difficult to constrain and yet to be fully understood.  
%===

At the innermost boundary ($r \lesssim 0.1$ au), the accretion of this gas is governed by the stellar magnetic field. In the standard magnetospheric accretion (MA) paradigm, the stellar magnetic field truncates the inner disk, channeling gas along magnetic field lines where it free-falls onto the star \citep{Koenigl1991ApJ...370L..39K, Shu1994ApJ...429..781S, Hartmann2016ARA&A..54..135H}. 
While ubiquitous in T Tauri stars, applying the MA paradigm to intermediate-mass Herbig Ae/Be stars is complex. Their typically weak magnetic fields ($B \lesssim 300$ G; \citealt{Wade2007MNRAS.376.1145W, Alecian2013MNRAS.429.1001A, Hubrig2015MNRAS.449L.118H}) combined with high accretion rates force the truncation radius inward, creating a highly compact, or `crushed,' magnetosphere\citep{Muzerolle2004ApJ...617..406M, Mendigutia2011A&A...535A..99M, Cauley2014ApJ...797..112C}. 
This violent boundary layer is simultaneously responsible for launching fast, magnetocentrifugal inner-disk winds and jets, intimately linking mass ejection to the accretion process \citep{Hartigan1995ApJ...452..736H,Ferreira2006A&A...453..785F}.

% 3. The Oxygen Line in this context (Setting up the anomaly) 

High-resolution optical spectroscopy provides a powerful probe to trace these extreme inner-disk kinematics. While permitted lines like H$\alpha$ and He I $\lambda$5876 trace the hot funnel flows and fast winds, the forbidden [O I] $\lambda$6300, 6363 \AA\ emission lines are critical diagnostics of the surrounding circumstellar geometry and slow outflows. In typical flared disks, [O I] emission originates from the UV-irradiated disk surface and shows symmetric, centrally peaked profiles indicative of stable Keplerian rotation \citep{Acke2005A&A...436..209A}. When participating in a slow photoevaporative or MHD disk wind, the [O I] profile exhibits a low-velocity blueshift ($-5$ to $-20$ km s$^{-1}$), as the optically thick disk midplane physically occults the receding, redshifted gas on the far side of the star \citep{Rigliaco2013ApJ...772...60R, Banzatti2019ApJ...870...76B}.
 Consequently, the absence of a wind-driven blueshift, combined with symmetric [O~I] emission centered at the stellar rest velocity, provides a unique kinematic signature. Rather than tracing an outflow, this profile isolates the gravitationally bound, stagnant gas reservoir pooling within the thermally cleared innermost cavity.

A benchmark laboratory for testing these accretion and ejection mechanisms is the Herbig Ae star AB Aurigae.
 AB Aurigae is a young, intermediate-mass Herbig Ae star ($M_ \approx 2.4 \ M_\odot$). While early studies using Hipparcos distances estimated its age at $4 \pm 1$ Myr \citep{DeWarf2003ApJ...590..357D}, modern pre-main-sequence evolutionary models incorporating the updated Gaia EDR3 distance ($d \approx 163$ pc) yield a slightly younger and broader age range of roughly $1$ to $4.4$ Myr \citep{Guzman2021A&A...650A.182G, Speedie2024Natur.633...58S}.
The system hosts a spectacular extended disk with a massive inner dust cavity spanning $\sim 70-120$ au \citep{Tang2017ApJ...840...32T} and exhibits prominent spiral arms linked to embedded protoplanet candidates \citep{Boccaletti2020A&A...637L...5B, Currie2022NatAs...6..751C}. Despite its evolved cavity, AB Aurigae maintains an active mass accretion rate of $\sim 10^{-7} M_\odot\ {\rm yr}^{-1}$. Recently, high-resolution ALMA observations by \citet{Speedie2025ApJ...981L..30S} revealed that the AB Aurigae system is currently undergoing a ``late-stage infall renovation," fueled by large-scale exo-disk streams from a gravitationally captured remnant cloud. While this massive external envelope successfully explains the replenishment of the outer disk, the mechanism by which this late-stage infall reaches the stellar surface and drives the high accretion rate remains observationally unconstrained.

To bridge the gap between the macro-scale ALMA observations and the micro-scale stellar accretion boundary, we conducted multi-epoch, high-resolution ($R \sim 107,000$) optical spectroscopy of AB Aurigae using the PARAS-2 spectrograph. By resolving the complex, day-to-day kinematics of the H$\alpha$, He I $\lambda$5876, and [O I] $\lambda$6300, 6363  and Na I D lines, we aim to map the physical structure of the innermost mass flow. In this article, we report the discovery of highly dynamic, clumpy magnetospheric funnel flows uniquely coupled with an anomalous, purely redshifted [O I] inflow. These kinematics provide the first direct optical evidence that the late-stage infall material successfully reaches the stellar environment to actively overload a crushed magnetosphere. Importantly, this reveals a novel diagnostic capability: the [O I] $\lambda$6300 forbidden line, which is traditionally established as a tracer of blueshifted disk winds and outflows \citep[e.g.,][]{Rigliaco2013ApJ...772...60R, Campbell2023ApJ...956...25C}, is unambiguously acting as a tracer of  stagnant gas reservoir pooling
within the thermally cleared innermost cavity.
The observations and data reduction are described in Section \ref{sec:obs}. The spectral analysis and kinematic results are presented in Section \ref{sec:results}. In Section \ref{sec:discussion}, we discuss the physical geometry of the crushed magnetosphere and the origin of the %anomalous 
oxygen  reservoir, followed by our conclusions in Section \ref{sec:conclusion}.

%%%%%%%%%%%SECTION 2 %%%%%%%%%%%%

\section{Observation} \label{sec:obs}

\subsection{The PARAS-2 Spectrograph and Instrumental Setup:} 
The high-resolution optical spectra of AB Aurigae were acquired using the PARAS-2 spectrograph attached to the 2.5 m telescope at the PRL Mount Abu observatory, Gurushikhar, India \citep{Chakraborty2024BSRSL..93...68C}. PARAS-2 is a state-of-the-art fiber-fed echelle spectrograph that works at a spectral resolution of $R \sim 107,000$ covering wavelengths from 380 nm to 690 nm in a single acquisition. To achieve very precise radial velocity measurements and to minimize the instrumental drift, the spectrograph kept in a customized thermally controlled vacuum chamber \citep{Chakraborty2018AJ....156....3C, Chakraborty2024BSRSL..93...68C}. The starlight and calibration light are transmitted from the telescope focal plane to the spectrograph via two octagonal fibers: the star fiber (A-fiber) and the calibration fiber (B-fiber) \citep{kevi2025}. However, for this work, all spectra were obtained using only the A-fiber; starlight was injected into the A-fiber during science exposures, while the calibration lamp light was injected into the same fiber during calibration exposures. 
 Because of the exceptional stability of the PARAS-2 spectrograph, the absolute instrumental drifts are of the order of a few m/s ($\leq$ 5 m/s)  for a typical night of 12 hours \citep{Chakraborty2024BSRSL..93...68C}. In the absence of simultaneous calibration spectra, such a minor drift would be insignificant for estimating any relevant shifts of the spectral lines, which are orders of magnitude larger ($\sim$ km/s).

\subsection{ Observations and calibration}
High-resolution optical spectra of the target star, AB Aurigae, were acquired over three consecutive nights on January 8, 9, and 10, 2026. Each science exposure was carried out with an integration time of 1800 s (30 minutes) to ensure a high signal-to-noise ratio (SNR) across the spectral orders. To avoid high atmospheric extinction and further optimize the SNR, the observations were scheduled such that the target was consistently observed at an airmass $\leq$1.2.
To accurately constrain the circumstellar emission toward the main target, each set of science observations was immediately followed by an exposure of the standard star HR 1544, with an integration time of about 900 s (15 minutes), observed at a similar airmass. HR 1544 is a rapidly rotating, featureless A1V standard star, making it an ideal reference to map and disentangle the Earth's atmospheric transmission. Dividing the target spectra by this standard allows for the precise removal of sharp telluric absorption lines, thereby isolating the uncontaminated emission profile of AB Aurigae.  

\subsection{Data Reduction and Telluric Correction}
The raw PARAS-2 echellograms were processed using the custom-built \texttt{PARAS-2 PIPELINE} software suite \citep{Baliwal2024}, which is based on the algorithms of \citet{Piskunov2002A&A...385.1095P} for the optimal extraction of cross-dispersed echelle spectra. The reduction process included master bias and dark subtraction, cosmic ray removal, scattered light correction before extracting the spectra. The precise wavelength solutions were derived using the Uranium lineist of \citet{Sharma2021JATIS...7c8005S}. More details of the data reduction pipeline and wavelength calibration can be found in \citet{Baliwal2024}. \\
Additionally, we performed polynomial fitting to individual spectral orders for the target source, manually masking intrinsic emission and absorption features to most precisely recover the  profile of the targeted spectra. To account for the Earth's rotational and orbital motion, a barycentric correction was applied to shift the velocities into the Solar System Barycenter frame using the algorithms given in \citet{Wright2014PASP..126..838W}. Finally, the systemic velocity ($v_{\rm rad} = +15.1$ km s$^{-1}$) of AB Aurigae was subtracted to shift the data into the stellar rest frame ( see Appendix A for detailed calculations and justifications). The spectrum of the standard star HR 1544 was then used to divide out the sharp telluric absorption lines, isolating the pure circumstellar emission profile of AB Aurigae

\section{Results} \label{sec:results}

%%%%%%%% all oxygen spectra

\begin{figure*}
    \centering
    \includegraphics[width=0.38\linewidth]{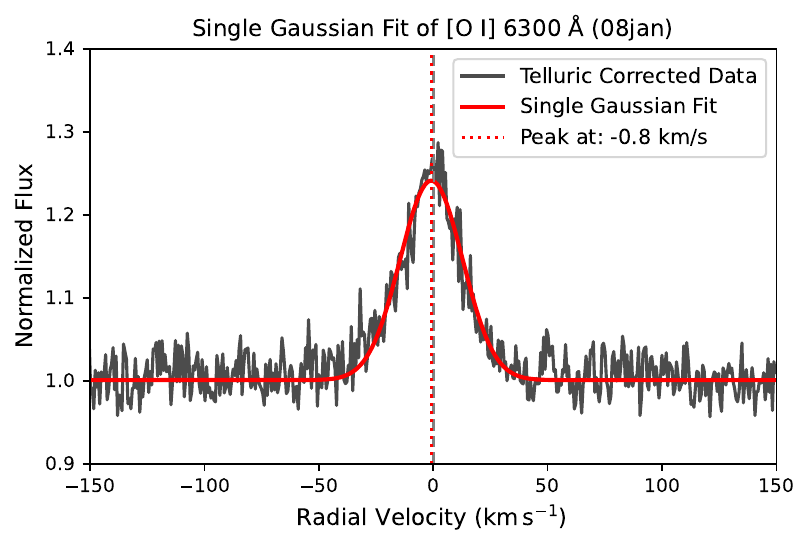}
    \includegraphics[width=0.38\linewidth]{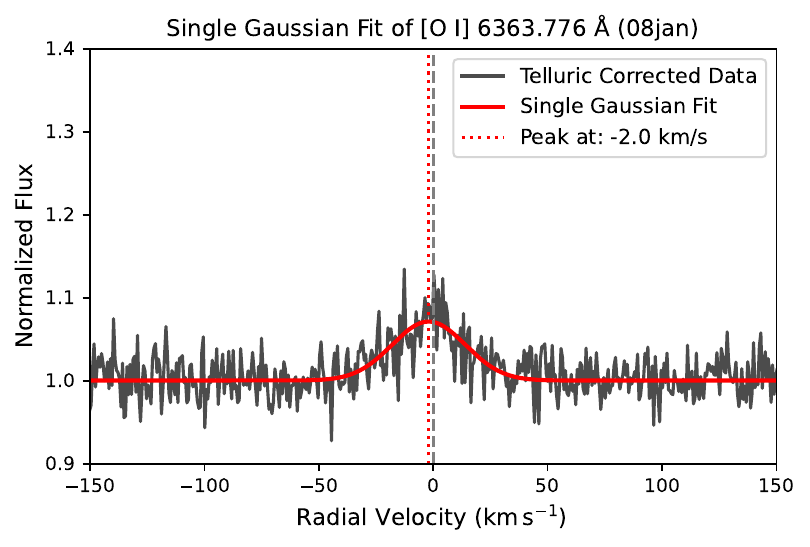}
    \includegraphics[width=0.38\linewidth]{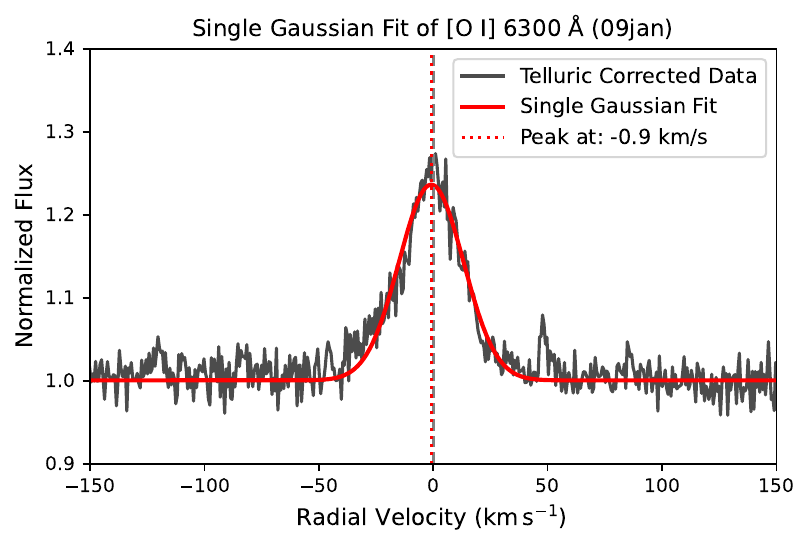}
    \includegraphics[width=0.38\linewidth]{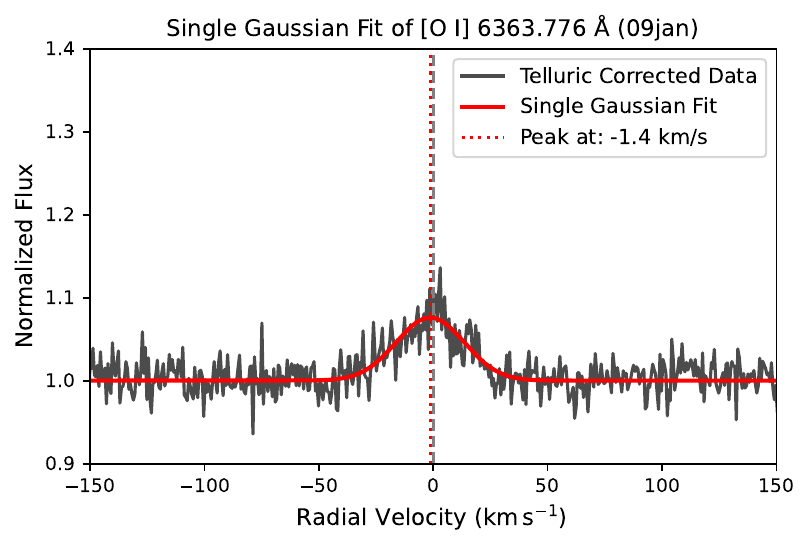}
    \includegraphics[width=0.38\linewidth]{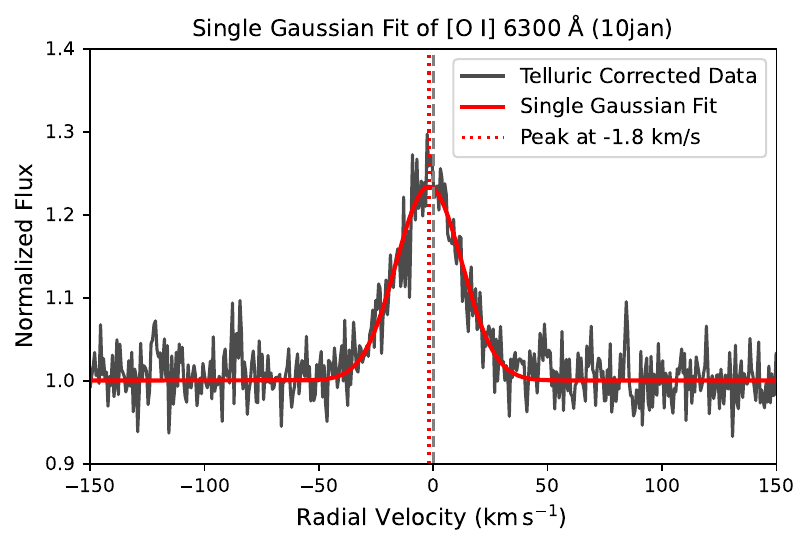}
    \includegraphics[width=0.38\linewidth]{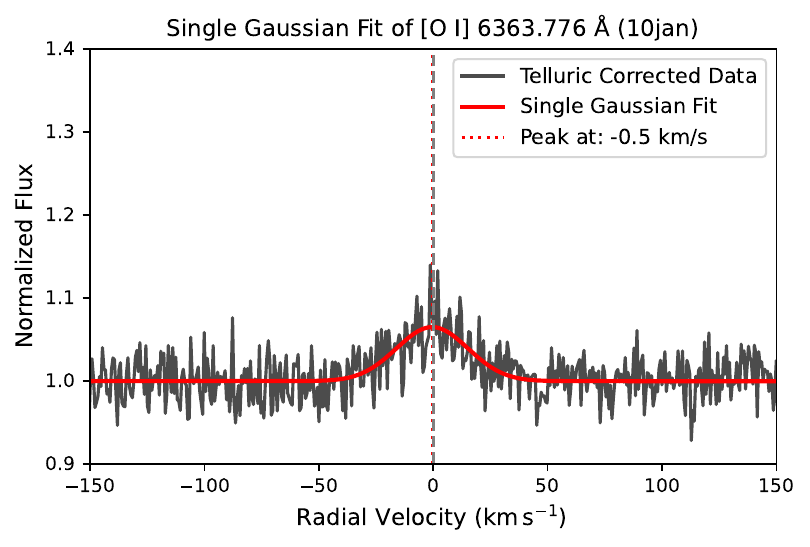}
    \caption{Telluric-corrected emission profiles of the forbidden oxygen [O~I] $\lambda$6300 (left), 6363 (right) \AA\ lines toward AB Aurigae, observed during January 8, 9, and 10, 2026. The spectra have been shifted to the stellar rest frame assuming a systemic radial velocity of $v_{\rm rad} = +15.1$ km s$^{-1}$. The vertical grey line denotes the stellar rest velocity at 0 km s$^{-1}$.}

    \label{fig:o1_all_epoch}
\end{figure*}

%%%%%%%% Updated table
\begin{deluxetable*}{l c c c c c}
\tablecaption{{ Single-Gaussian Fitting Parameters for the [O~I] Emission and He~I Absorption Lines toward AB Aurigae} \label{tab:oi_hei_fitting_results}}
\tablewidth{0pt}
\tablehead{
\colhead{Observation Date} &
\colhead{Spectral Line} &
\colhead{Peak Velocity ($v_c$)} &
\colhead{FWHM\tablenotemark{a}} &
\colhead{Gaussian Area, $W_v$\tablenotemark{b}} &
\colhead{EW\tablenotemark{c}} \\
\colhead{(2026)} &
\colhead{($\AA$)} &
\colhead{(km s$^{-1}$)} &
\colhead{(km s$^{-1}$)} &
\colhead{(km s$^{-1}$)} &
\colhead{($\AA$)}
}
\startdata
January 8  & [O~I] 6300.3 & $-0.83 \pm 0.37$ & $32.10 \pm 0.86$ & $8.210 \pm 0.292$ & $-0.173 \pm 0.006$ \\
           & [O~I] 6363.8 & $-1.99 \pm 1.31$ & $37.69 \pm 3.09$ & $2.851 \pm 0.309$ & $-0.061 \pm 0.007$ \\
           & He~I 5876    & $25.83 \pm 1.15$ & $166.02 \pm 4.59$ & $24.131 \pm 0.818$ & $+0.473 \pm 0.016$ \\
\hline
January 9  & [O~I] 6300.3 & $-0.87 \pm 0.32$ & $33.15 \pm 0.74$ & $8.321 \pm 0.247$ & $-0.175 \pm 0.005$ \\
           & [O~I] 6363.8 & $-1.35 \pm 0.92$ & $35.09 \pm 2.18$ & $2.836 \pm 0.233$ & $-0.060 \pm 0.005$ \\
           & He~I 5876    & $52.73 \pm 1.63$ & $161.00 \pm 5.64$ & $15.783 \pm 0.690$ & $+0.309 \pm 0.014$ \\
\hline
January 10 & [O~I] 6300.3 & $-1.83 \pm 0.43$ & $33.85 \pm 1.02$ & $8.408 \pm 0.336$ & $-0.177 \pm 0.007$ \\
           & [O~I] 6363.8 & $-0.45 \pm 1.58$ & $36.80 \pm 3.73$ & $2.552 \pm 0.341$ & $-0.054 \pm 0.007$ \\
           & He~I 5876    & $27.72 \pm 1.23$ & $167.73 \pm 4.89$ & $26.617 \pm 0.952$ & $+0.522 \pm 0.019$ \\
\enddata
\tablenotetext{a}{Line width is the Full Width at Half Maximum (FWHM) of the fitted Gaussian profile.}
\tablenotetext{b}{$W_v = A \sigma \sqrt{2\pi}$ is the integrated area of the fitted Gaussian profile in velocity units.}
\tablenotetext{c}{Classical equivalent width derived from the Gaussian area using $EW_\lambda = (\lambda_0/c)\,W_v$. Emission lines are reported with negative EWs, while absorption lines are positive. The [O~I] $\lambda6300$ EW is the quantity used in Appendix~\ref{sec:appendix_oi_temp} to derive a typical line flux.}
\end{deluxetable*}
%%%%%%%%%

\subsection{Detection of [O~I] Spectra } 

After confirming the intrinsic circumstellar nature of the forbidden oxygen emission toward AB Aurigae (see Appendix~\ref{sec:appendix_oi_temp}), we analyzed the telluric-corrected spectra for each individual observing epoch. In addition to the primary [O~I] $\lambda 6300$ transition, we successfully detected the weaker satellite transition at [O~I] $\lambda 6363$. %, whose flux ratio relative to $\lambda 6300$ provides an independent constraint on the line optical depth (see Appendix A). 
The final telluric-corrected [O~I] profiles across all three observing epochs are presented in Figure~\ref{fig:o1_all_epoch}.

In typical Herbig Ae/Be systems with passive, flared circumstellar disks, this non-thermal emission line (driven by UV photodissociation of OH and H$_2$O molecules) is expected to be symmetrically centered at the stellar rest velocity %\citep[$v_{\rm rad} = +8.9$ km s$^{-1}$,][]{Gontcharov2006AstL...32..759G}
if it originates from purely rotating Keplerian gas \citep{Acke2005A&A...436..209A}. Alternatively, it may exhibit a blueshifted low-velocity component (LVC) if tracing an outward-blowing photoevaporative or magnetohydrodynamic (MHD) disk wind \citep{Pascucci2020ApJ...903...78P}. 

To assess these physical processes, we performed multiple component spectral fits and found that a single-component Gaussian profile best fits both of the forbidden oxygen emission features. The results of this Gaussian fitting are presented in Table~\ref{tab:oi_hei_fitting_results}.
 The fitting results (Table~\ref{tab:oi_hei_fitting_results}) reveal a single-peaked profile centered at $\sim -0.8$ to -$1.8$ km s$^{-1}$ in the stellar rest frame. Given the instrumental precision and systemic velocity ($v_{\rm rad} = +15.1$ km s$^{-1}$, see Appendix) uncertainties, this emission is effectively centered at $v \approx 0$ km s$^{-1}$. In particular, we do not detect the characteristic $5$ to $20$ km s$^{-1}$ blueshift typically associated with photoevaporative or MHD disk winds \citep{Rigliaco2013ApJ...772...60R}. The strict absence of this blueshift, combined with the symmetric broadening of the line ($\sim 35$ km s$^{-1}$), implies that the gas is not participating in a vertical outflow, but rather resides in a bound, rotating configuration \citep{Acke2005A&A...436..209A}.

Additionally, the integrated [O~I] $\lambda 6300 / \lambda 6363$ flux ratio remains consistently $\sim 3.0$ across all epochs, closely matching their Einstein $A$ transition probabilities; this also implies that the emission originates in an optically thin regime \citep{Acke2005A&A...436..209A}.
Based on the normalized equivalent widths and the model-derived continuum flux at the exact line wavelengths, we derive an absolute  line luminosity of $L_{6300} \approx 6.62 \times 10^{-4} \, L_\odot$ (see Appendix~\ref{sec:appendix_oi_temp}).  
We note that \citet{Acke2005A&A...436..209A} previously detected [O~I] $\lambda$6300 emission toward AB Aurigae with similar $L_{6300}$, and reported a centroid redshift of $+15$ km s$^{-1}$ in the stellar rest frame based on an adopted systemic velocity of $v_{\rm sys} = 9$ km s$^{-1}$. If we correct their absolute line position using our highly precise ALMA-anchored systemic velocity ($v_{\rm sys} = 15.1$ km s$^{-1}$), their residual rest-frame offset reduces to approximately $+9$ km s$^{-1}$. This residual shift is entirely consistent with zero given the calibration uncertainties associated with their lower spectral resolution ($R \approx 30{,}000$--$65{,}000$, corresponding to resolution elements of $5$--$10$ km s$^{-1}$). Furthermore, our extreme-resolution PARAS-2 data ($R \approx 107{,}000$) resolve the full kinematic profile, revealing a true FWHM of $\sim 35$ km s$^{-1}$ compared to the narrower $\sim 20$ km s$^{-1}$ as  reported by \citep{Acke2005A&A...436..209A}. By pairing our extreme spectral resolution observations with the ALMA-derived disk kinematic center, we resolve the apparent redshift discrepancy present in the earlier literature. Our observations thus provide definitive kinematic evidence that the [O~I] emission in AB Aurigae is definitively centered at the stellar rest velocity, tracing a gravitationally bound Keplerian reservoir. 

We do not detect the higher-excitation [O~I] $\lambda$5577 transition. Using the local continuum noise level, the $3\sigma$ upper limit on the $\lambda$5577 equivalent width yields a strict line luminosity ratio of {\bf $L_{6300}/L_{5577} > 4.8$.} Assuming Local Thermodynamic Equilibrium (LTE) in a high-density regime where collisional de-excitation dominates, this ratio places a strict upper limit on the gas excitation temperature of {\bf $T \lesssim 3829$ K} (Appendix~\ref{sec:appendix_oi_temp}). This relatively low temperature limit confirms that the [O~I] emission does not originate in an ultra-hot, highly ionized accretion shock at the stellar surface.  Instead, it traces a cooler, gravitationally bound, stagnant gas reservoir pooling within the innermost cavity prior to magnetospheric accretion.

%%%%%%%%%%--------------

\begin{figure*}
    \centering
    \includegraphics[width=0.3\linewidth]{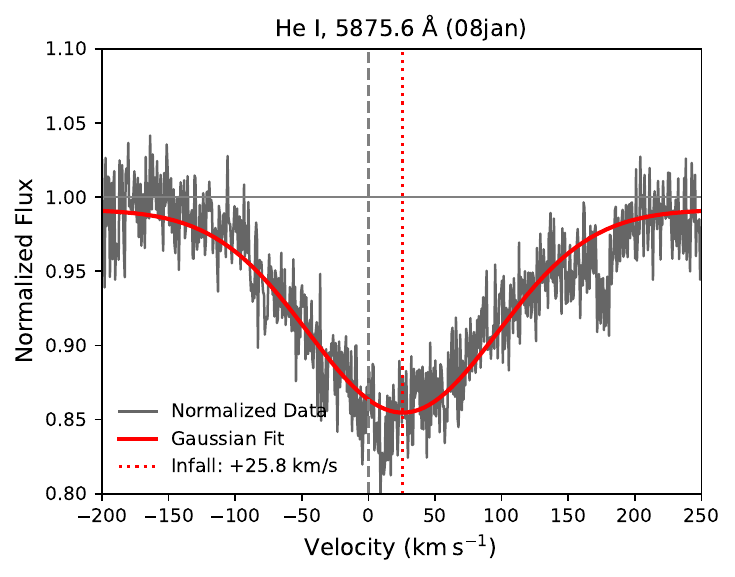}
    \includegraphics[width=0.3\linewidth]{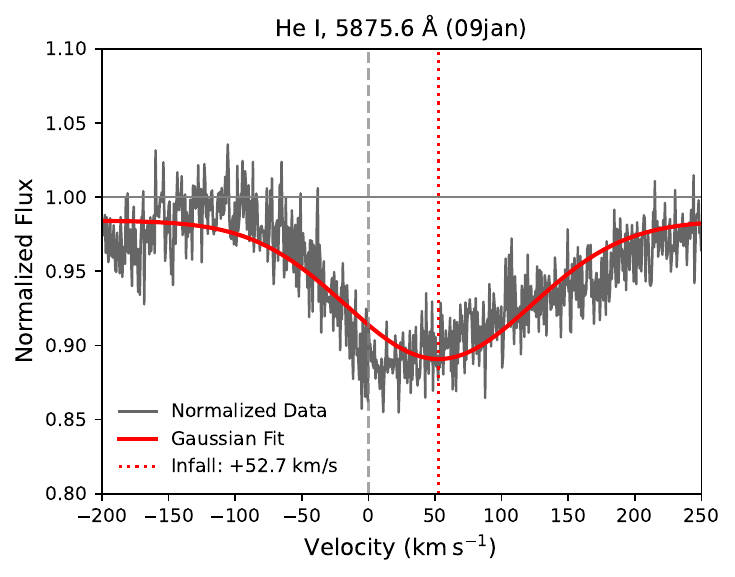}
    \includegraphics[width=0.3\linewidth]{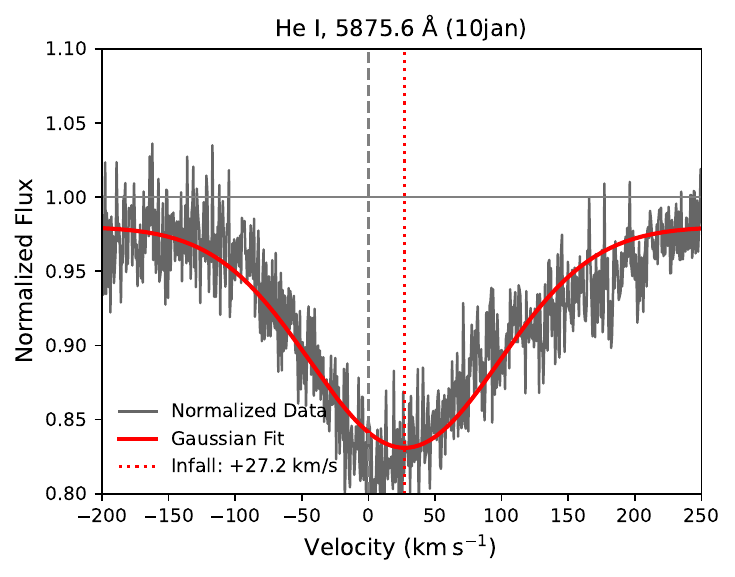}
%====================H-alpha after this 
    \includegraphics[width=0.45\linewidth]{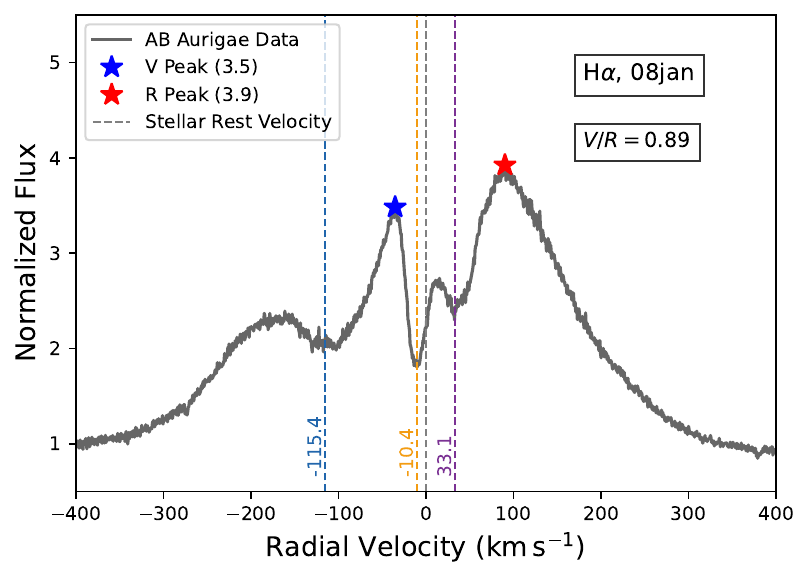}
    \includegraphics[width=0.45\linewidth]{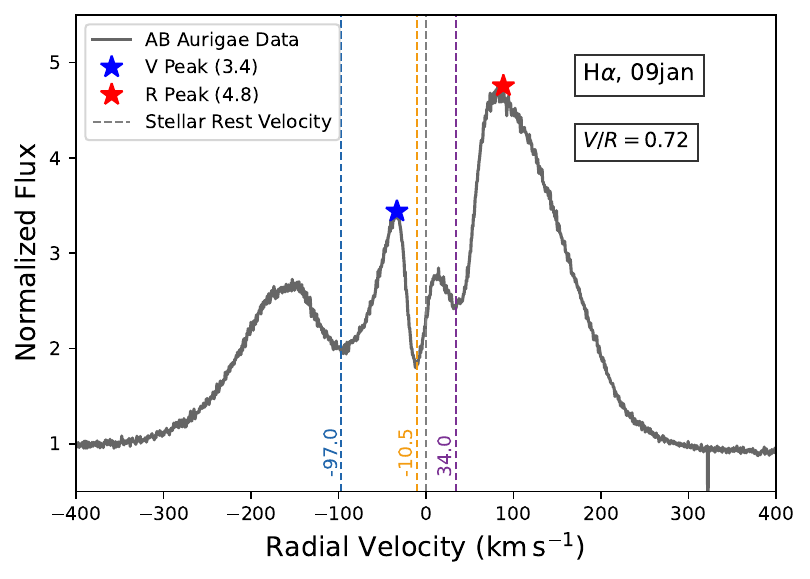}
    \includegraphics[width=0.45\linewidth]{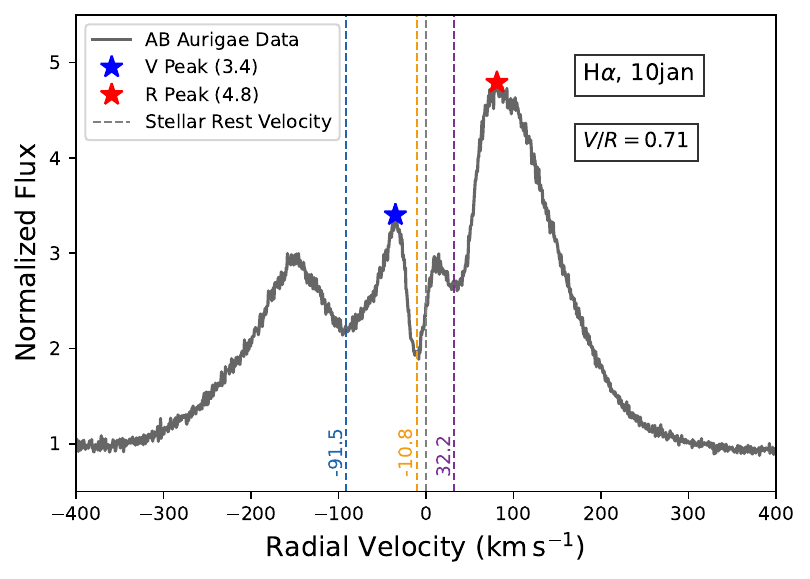}
    
    \caption{He I and H$\alpha$ profile accross three epochs. }
    \label{fig:He1_Ha_combined}
\end{figure*}

%%%%%%%%%Na time series
\begin{figure*}[ht!]
    \centering
    \includegraphics[width=0.7\linewidth]{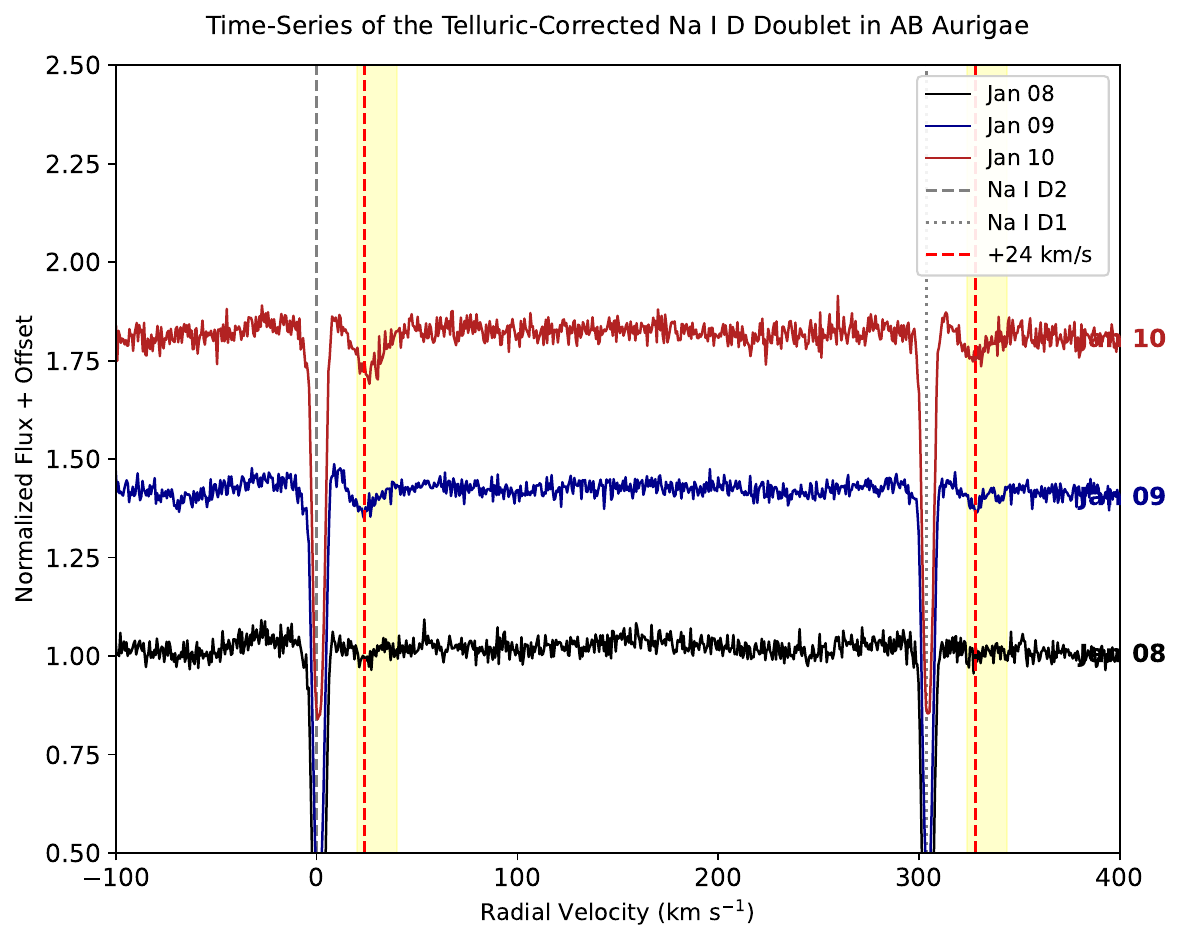}
    \caption{Time-series evolution of the telluric-corrected Na I D doublet ($\lambda 5889, 5895$) toward AB Aurigae across the three observing epochs (January 8, 9, and 10, 2026). The spectra are normalized and shifted vertically for clarity. The rest velocities of the D2 and D1 transitions are marked by vertical grey lines. Both transitions exhibit a highly variable, broad redshifted absorption trough centered at $\approx +24$ km s$^{-1}$; indicated by vertical red dashed lines and yellow shaded regions of total width of 20~km s$^{-1}$. }
    \label{fig:NaI}
\end{figure*}

%%%%%%%%%%UPdated table 

\begin{deluxetable*}{lcccccccc}
\tablecaption{ { H$\alpha$ Emission-Line Profile Parameters and Accretion Rates for AB Aurigae}\label{tab:halpha_params}}
\tablewidth{0pt}
\tablehead{
    \colhead{Date} &
    \colhead{$V/R$} &
    \colhead{$V_{\rm peak}$\tablenotemark{a}} &
    \colhead{$R_{\rm peak}$\tablenotemark{a}} &
    \colhead{Blue Dip\tablenotemark{b}} &
    \colhead{Central Dip\tablenotemark{b}} &
    \colhead{Red Dip\tablenotemark{b}} &
    \colhead{Tot.\ EW\tablenotemark{c}} &
    \colhead{$\dot{M}_{\rm acc}$\tablenotemark{d}} \\
    \colhead{(2026)} &
    \colhead{} &
    \colhead{(km s$^{-1}$)} &
    \colhead{(km s$^{-1}$)} &
    \colhead{(km s$^{-1}$)} &
    \colhead{(km s$^{-1}$)} &
    \colhead{(km s$^{-1}$)} &
    \colhead{(\AA)} &
    \colhead{($M_{\odot}$ yr$^{-1}$)}
}
\startdata
Jan 8  & 0.89 & $-35.4$ & $90.1$ & $-115.4 \pm 0.4$ & $-10.4 \pm 0.1$ & $33.1 \pm 0.4$ & $-17.3 \pm 0.9$ & $(3.7 \pm 2.9)\times10^{-7}$ \\
Jan 9  & 0.72 & $-33.4$ & $88.3$ & $-97.0 \pm 0.2$  & $-10.5 \pm 0.1$ & $34.0 \pm 0.4$ & $-18.3 \pm 0.9$ & $(3.9 \pm 3.1)\times10^{-7}$ \\
Jan 10 & 0.71 & $-35.1$ & $80.9$ & $-91.5 \pm 0.3$  & $-10.8 \pm 0.2$ & $32.2 \pm 0.4$ & $-18.8 \pm 0.9$ & $(4.0 \pm 3.2)\times10^{-7}$ \\
\enddata
\tablenotetext{a}{$V_{\rm peak}$ and $R_{\rm peak}$ denote the velocities of the blue and red H$\alpha$ emission peaks, respectively, measured in the stellar rest frame.}
\tablenotetext{b}{The Blue Dip, Central Dip, and Red Dip columns list the centroid velocities of the three absorption minima in the H$\alpha$ profile, measured in the stellar rest frame.}
\tablenotetext{c}{Total equivalent width of the H$\alpha$ line in \AA. Negative values indicate net emission, following standard spectroscopic convention.}
\tablenotetext{d}{Accretion rates derived from the H$\alpha$ line luminosity using the empirical calibration of \citet{Fairlamb2017MNRAS.464.4721F}. Uncertainties include both the propagated measurement error and the intrinsic scatter of the calibration relation ($\sim$0.25 dex).}
\tablecomments{All velocities have been corrected to the adopted stellar systemic velocity of 15.1 km s$^{-1}$. The $V/R$ ratio is the ratio of the blue to red H$\alpha$ emission-peak intensities.}
\end{deluxetable*}

%%%%%%%%%%%%%%%%%%%%%%%%%%%%%%%%

\subsection{ He I and Na I D: Tracer of Magnetospheric Accretion}
%\subsection{He I $\lambda$5876: Tracing the Crushed Magnetosphere and Rotational Modulation}
\label{sec:he1}

To constrain the kinematics of the inner disk and validate the accretion-driven origin of the [O~I] emission, we analyzed the He~I $\lambda$5876 line (He~I $\mathrm{D_3}$). As the excitation potential of the He~I transitions is exceptionally high, this line exclusively traces the hottest and most energetic regions of the star-disk interface. The He I $\lambda$5876 line is a well-established diagnostic of magnetospheric accretion in both classical T Tauri stars (CTTSs), and Herbig Ae/Be stars \citep{Muzerolle2004ApJ...617..406M, Fairlamb2017MNRAS.464.4721F}. Specifically, it  probes the hot, dense post-shock cooling region at the footpoints of the magnetospheric funnel flow. 

Our He I $\lambda$5876 line observations (Figure~\ref{fig:He1_Ha_combined}) show broad, purely redshifted absorption troughs in the line profile across all epochs. The emission feature typically observed in an inverse P-Cygni profile is not detected with enough S/N,  likely as the post-shock region is projected mostly transverse to the line of sight. To quantitatively analyze the absorption troughs, a single Gaussian fitting is performed. The absorption depths ($A \approx 0.09-0.15$) imply partially optically thin conditions in the accretion funnel. \citep{Muzerolle2004ApJ...617..406M}.
The He I line is unaffected by telluric contamination at 5876 \AA, which make these measurements robust diagnostics of the instantaneous accretion state.

 The fitting results as noted in Table~\ref{tab:oi_hei_fitting_results} reveal redshifted absorption centroids ($v_c$) that vary significantly on daily timescales, shifting from $+25.8 \pm 1.1$ km s$^{-1}$ (Jan 8) to $+52.7 \pm 1.6$ km s$^{-1}$ (Jan 9), and returning to $+27.7 \pm 1.2$ km s$^{-1}$ (Jan 10). In parallel, the normalized equivalent width decreases from $24.13 \pm 0.82$ to $15.78 \pm 0.69$ and then rises again to $26.62 \pm 0.95$, closely tracking the change in absorption geometry. Therefore, the significant night-to-night variation ($\Delta v_c \sim 25$ km s$^{-1}$) indicates a highly non-axisymmetric, time-variable accretion flow, consistent with a changing magnetospheric funnel geometry and a variable truncation radius  (see the schematic diagram in Fig.~\ref{fig:schematic_diagram_AB_aurigae}).

Notably, the lines are massively broadened with a Full Width at Half Maximum (FWHM) of $ \approx 161-168$ km s$^{-1}$. In the magnetospheric accretion (MA) paradigm, the maximum redshifted velocity ($V_{\rm red, max}$) of the absorption profile records the terminal free-fall velocity ($v_{\rm ff}$) of the accreting gas as it impacts the star. From our profiles, the maximum redshifted extent of the absorption (measured at 10\% below the continuum) reaches $V_{\rm red, max} \approx 130$ km s$^{-1}$. For AB Aurigae ($M_* = 2.4 M_\odot$, $R_* = 2.62 R_\odot$), the stellar surface escape velocity is $v_{\rm esc} \approx 600$ km s$^{-1}$. The observation that the terminal infall velocity is a mere fraction of the escape velocity ($V_{\rm red, max} / v_{\rm esc} \approx 0.21$) physically demands that the gas falls from a very short distance.  Assuming the reported upper-limit magnetic field strength of $B \lesssim 300\,\mathrm{G}$, the derived magnetospheric truncation radius is $R_{\rm mag} \approx 1.2 R_*$. (see Appendix~\ref{sec:appendix_kinematics}).
A free-fall trajectory originating from this crushed truncation radius closely reproduces the observed $130$ km s$^{-1}$ terminal velocity.  Note that a smaller true magnetic field would imply an even smaller truncation radius and, consequently, a smaller free-fall velocity at the stellar surface.

In addition to this, the rapid night-to-night variation of the absorption centroid ($\Delta v_c \sim 25$ km s$^{-1}$) occurs on timescales consistent with the stellar rotation period of AB Aurigae ($P_{\rm rot} \approx 34$ hours, assuming $v \sin i = 116$ km s$^{-1}$). This daily modulation strongly suggests that the inner magnetosphere is not a steady, axisymmetric structure.

The Na~I~D doublet ($\lambda 5889, 5895$) is a well-established tracer of the cooler, extended regions of magnetospheric funnel flows in young stellar objects \citep[e.g.,][]{Mora2004A&A...419..225M, Muzerolle2004ApJ...617..406M}. These transitions complement our high-excitation He~I $\lambda$5876 observations, which trace the localized hot accretion shock at the stellar surface.

Historical spectral atlases of AB Aurigae \citep{Bouret1998A&A...340..163B, Bohm1993A&AS..101..629B} show that the Na~I~D profiles are intrinsically dynamic, with substantial morphological changes on timescales of weeks to months. Our multi-epoch PARAS-2 observations demonstrate that this variability extends down to daily timescales in the inner accretion environment. While the intense Na~I~D absorption components, likely associated with interstellar medium, remain kinematically stationary over our three-night window, the redshifted inverse P-Cygni absorption features at $\approx +24$ km s$^{-1}$  are highly variable. On January 8, this component is weak or absent, but it rapidly develops and deepens over the next 48 hours, becoming prominent on January 9 and 10 (Figure~\ref{fig:NaI}). This evolution is the classic signature of changing magnetospheric accretion and indicates that a dense, discrete clump of cooler neutral gas passed through the accretion funnel during our observing window, consistent with the clumpy, episodic mass-loading inferred from the H$\alpha$ (see next section) and He~I kinematics.

\subsection{Kinematics and Temporal Variability of the H$\alpha$ Profile} \label{subsec:halpha}

%%modified section descriptions 
The H$\alpha$ emission line is a powerful diagnostic tool for probing the immediate circumstellar environment, simultaneously tracing both mass accretion and outflow activities in Herbig Ae/Be systems \citep{Muzerolle2001ApJ...550..944M, Mendigutia2011A&A...535A..99M}. Figure~\ref{fig:He1_Ha_combined} shows the H$\alpha$ profiles of AB Aurigae across our three observing epochs (January 8, 9, and 10, 2026). The spectra exhibit a complex composite P-Cygni profile - highly broadened morphology characterized by an intense emission peak, a prominent inverse P-Cygni profile showing redshifted absorption, and a broad, variable blueshifted absorption dip. To find the positions of the dips, we fit  reverse Gaussians to extract the peak velocity. To check the variation of blue and red peaks over the epochs, we have calculated their ratios (V/R); all these results are noted in the Table~\ref{tab:halpha_params}. Additionally, we have calculated the accretion from  the effective width (EW) of H$\alpha$ emission and using the method as explained by \citet{Fairlamb2015MNRAS.453..976F} (see Appendix~\ref{sec:appendix_macc}).

 We resolve the redshifted absorption (the ``Red Dip'') at  relatively stable velocities of $+32$ to $+34$ km s$^{-1}$ across all three observing epochs (Table~\ref{tab:halpha_params}). Simultaneously, the H$\alpha$ profiles exhibit a highly dynamic two-component blueshifted absorption, which is a classic signature of mass outflow obscuring the underlying emission \citep{Kurosawa2006MNRAS.370..580K, Herbig2007AJ....133.2679H}. We detect a high-velocity ``Blue Dip'' that rapidly decelerates from $-115.4$ km s$^{-1}$ down to $-91.5$ km s$^{-1}$ over three days.  It can be noted from Table~\ref{tab:halpha_params} that the accretion rate ($\dot{M}_{\rm acc}$) shows only modest epoch-to-epoch variation, with values of order $4\times10^{-7}\,M_\odot\,{\rm yr}^{-1}$ within the quoted uncertainties. Consequently, the H$\alpha$ profile exhibits a highly variable blue-shifted absorption dip, accompanied by a decreasing $V/R$ ratio across the epochs. This behavior is consistent with the theoretical and observational expectation of episodic or clumpy accretion events \citep{Calvet2004AJ....128.1294C}. In addition to the high-velocity wind component, we also resolve a stable, low-velocity ``Central Dip'' fixed at $-10.4$ to $-10.8$ km s$^{-1}$. This stationary feature likely traces the base of a steady, low-velocity disk wind driven by photoevaporation or large-scale MHD processes further out in the disk \citep{Alexander2014prpl.conf..475A, Pascucci2020ApJ...903...78P}.

\section{Discussion} \label{sec:discussion}

%%%%%final modified section

\subsection{AB Aurigae in the Context of Previous H$\alpha$ Studies}

 AB Aurigae has long been recognized as a highly variable Herbig Ae system whose optical line profiles cannot be understood within a steady, spherically symmetric outflow picture. Earlier spectroscopic atlases and monitoring campaigns \citep[e.g.,][]{Bohm1993A&AS..101..629B, Bouret1998A&A...340..163B, Bouret2000A&A...359.1011B} showed that the H$\alpha$, He~I, and Na~I~D profiles of the star change substantially from epoch to epoch, both in morphology and in absorption/emission strength. Our 2026 PARAS-2 spectra must therefore be interpreted not as an isolated snapshot, but as part of a dynamically evolving circumstellar environment.

A crucial point of comparison is the optical interferometric study of \citet{Rousselet2010A&A...516L...1R}, who spatially resolved the H$\alpha$-emitting region of AB Aurigae on sub-au scales. Their analysis explicitly ruled out a spherical stellar wind, instead favoring an extended, strongly non-spherical geometry consistent with magneto-centrifugal acceleration at the star--disk interface. Our spectroscopic interpretation is fully consistent with this picture: the complex H$\alpha$ morphology we observe is naturally explained by multiple kinematic components embedded in a highly structured accretion--ejection environment.

The fast blueshifted H$\alpha$ absorption component (the ``Blue Dip'') spans velocities from $\sim -115$ to $-92$ km s$^{-1}$ in our adopted stellar rest frame. This deep absorption represents a  P-Cygni profile and traces the accelerating inner outflow. While its overall magnitude is consistent with the historical low-resolution atlases of AB Aurigae \citep{Bouret1998A&A...340..163B}, our extreme-resolution, multi-epoch data reveal severe short-term kinematic fluctuations. Over just 48 hours, the peak absorption velocity of the wind decelerates by nearly $23$ km s$^{-1}$, indicating that the inner accretion--ejection process is not in a steady state but is instead strongly perturbed by episodic mass loading.

In direct kinematic contrast to this fast wind, the narrow low-velocity H$\alpha$ absorption component (the ``Central Dip'') remains stable near $-10.5$ km s$^{-1}$. For the nearly face-on inclination of AB Aurigae, this slow component is consistent with the base of a low-velocity disk wind driven by photoevaporation or large-scale MHD processes in the outer disk surface layers \citep[e.g.,][]{Alexander2014prpl.conf..475A, Ercolano2017RSOS....470114E}. The key point is that this stable low-velocity absorption is kinematically distinct from the faster wind traced by the Blue Dip, implies that we can disentangle multiple flow components within the H$\alpha$ profile.

While the superposition of fast inner winds and slower outer disk winds has been well documented in young stellar objects \citep[e.g.,][]{Kurosawa2006MNRAS.370..580K, Banzatti2019ApJ...870...76B}, observing both simultaneously in the H$\alpha$ profile of a Herbig Ae star is exceptionally challenging. The remarkable temporal stability of the low-velocity component, together with the violent day-to-day variability of the high-velocity wind, highlights the complexity of the AB Aurigae inner environment and provides the observational basis for the more detailed accretion and reservoir interpretation developed in the following subsections.

\subsection{The Mass Accretion Rate and the Late-Stage Infall Supply}

A fundamental puzzle in transition disks is how central stars maintain vigorous mass accretion despite massive, planet-carved cavities that theoretically throttle radial gas transport \citep[e.g.,][]{Mendigutia2012A&A...543A..59M}. Using the empirical H$\alpha$ luminosity relationships of \citet{Fairlamb2017MNRAS.464.4721F} and standard magnetospheric energy balance equations (detailed in Appendix~\ref{sec:appendix_macc}), we derive a highly active mass accretion rate that remains broadly consistent across our three epochs, with values ranging from $\dot{M}_{\rm acc} \approx (3.7 \pm 2.9)\times 10^{-7}$ to $(4.0 \pm 3.2)\times 10^{-7}\,M_\odot\,{\rm yr}^{-1}$. While this is slightly higher than historical estimates based on older distance calibrations \citep[e.g.,][]{Garcia2006A&A...459..837G, Salyk2013ApJ...769...21S}, it confirms that AB Aurigae is accreting at a rate comparable to, or even exceeding, many full (non-cavity) Herbig Ae disks.

Sustaining this massive mass transfer requires continuous replenishment of the inner mass reservoir. High-resolution ALMA observations recently mapped large-scale ``late-stage infall'' streamers actively feeding fresh gas into the outer disk at $r \sim 200$ au \citep{Speedie2025ApJ...981L..30S}. While ALMA traced the outer envelope of this infall, the ultimate kinematic fate of this material deep within the cavity has remained unconstrained. Our optical kinematics provide the direct inner-boundary ($r \lesssim 0.1$ au) counterpart to this large-scale mass transport, tracing the freshly supplied material as it interacts with the central star.

Further constraints on the physical state of this accreting gas come from the Na~I~D doublet ($\lambda 5889, 5895$), a well-established tracer of the cooler, extended regions of magnetospheric funnel flows in young stellar objects \citep[e.g.,][]{Mora2004A&A...419..225M, Muzerolle2004ApJ...617..406M}. As noted earlier, our Na~I profiles exhibit broad, redshifted inverse P-Cygni absorption troughs whose depth tracks the instantaneous accretion state. The observed D2/D1 absorption depth ratio is close to 2:1, suggesting that the sodium-bearing material is not  saturated. Furthermore, the survival of neutral sodium within this high-velocity flow requires gas temperatures of $T \lesssim 6000$,K, as hotter environments would result in complete ionization \citep{Muzerolle2004ApJ...617..406M}.

While the sudden deepening of the Na~I absorption could theoretically be produced by the infall of a large ($\sim 100$ km) planetesimal or boulder \citep[e.g.,][]{Chakraborty2004ApJ...606L..69C}, continuous solid infall is highly unlikely in AB Aurigae because the massive planet-carved gap efficiently halts the inward radial drift of large solids. Instead, the correlated behavior of the Na~I, He~I, and H$\alpha$ features is entirely consistent with a self-shielded, episodic gas accretion scenario (see Appendix \ref{sec:appendix_na}). This confirms that the observed inner-disk kinematics are driven by continuous, clumpy gas flows, directly linking the stellar magnetosphere to the macro-scale late-stage infall.

\subsection{The Crushed Magnetosphere and Funnel-Flow Kinematics}

AB Aurigae possesses a weak measured magnetic field with an upper limit of $B \approx 300$~G \citep[e.g.,][]{Wade2007MNRAS.376.1145W, Alecian2013MNRAS.429.1001A}. The immense ram pressure from the intense accretion rate violently compresses this field, creating a \emph{crushed} magnetosphere with a highly compact truncation radius of $R_{\rm mag} \approx 1.2\,R_\star$ (see Appendix~\ref{sec:appendix_kinematics}). Gas coupled to the magnetic field at this boundary rotates at the local Keplerian velocity, which naturally explains the extreme broadening observed in the H$\alpha$ emission wings ($\Delta W_{10\%} \approx 500$ km s$^{-1}$).

The He~I $\lambda5876$ transition provides the most direct probe of the hottest post-shock gas at the magnetospheric footpoints on the stellar surface \citep{Muzerolle2004ApJ...617..406M, Fairlamb2017MNRAS.464.4721F}. In our spectra, the line is dominated by broad, redshifted absorption troughs whose centroid velocity changes substantially from night to night. When the nearly face-on viewing geometry of AB Aurigae ($i \sim 22^\circ$; \citealp{Corder2005ApJ...622L.133C}) is taken into account, the observed maximum redshifted extent of the absorption ($V_{\rm red,max} \approx 130$--$140$ km s$^{-1}$) is consistent with gas falling from a very short distance, as expected for a compact truncation radius near $1.2\,R_\star$.

The same line also reveals rapid morphological changes in the absorption centroid, shifting from $v_c \approx 25.8$ km s$^{-1}$ on January 8 to $52.7$ km s$^{-1}$ on January 9 and then back to $27.7$ km s$^{-1}$ on January 10. This $> 25$ km s$^{-1}$ swing on daily timescales indicates that the funnel flows are highly non-axisymmetric and clumpy, rather than steady streams. Such rapid variability is naturally expected if the inner magnetosphere is continually perturbed by unsteady mass loading from the surrounding accretion reservoir.

Taken together, the H$\alpha$ wings and the He~I absorption strongly support a scenario in which the stellar magnetosphere is being compressed to a very small radius and intermittently overloaded by dense accretion streams. In this picture, the H$\alpha$ profile traces the large-scale kinematic response of the inner accretion--ejection flow, while He~I directly probes the final, hottest stage of the infalling gas before it impacts the stellar surface.

\subsection{The Origin of [O~I] Emission: Bound Keplerian Gas at 1 au}
\label{sec:oi_discussion}

In young stellar objects, the forbidden oxygen emission lines such as [O~I] $\lambda6300$ are classical diagnostics of the circumstellar gas geometry. Historically, this emission is divided into a High-Velocity Component (HVC) tracing fast, collimated jets and a Low-Velocity Component (LVC) tracing slower winds or bound disk gas \citep{Hartigan1995ApJ...452..736H}. High-resolution surveys have shown that the [O~I] LVC cannot generally be reproduced by purely thermal emission from X-ray- or EUV-heated layers, because such models fail to explain the observed low [O~I] $\lambda6300/\lambda5577$ line ratios \citep[e.g.,][]{Rigliaco2013ApJ...772...60R, Pascucci2020ApJ...903...78P}. Instead, the LVC is typically linked to non-thermal photodissociation of OH and H$_2$O molecules by stellar far-ultraviolet photons at the disk surface.

As discussed by \citet{Rigliaco2013ApJ...772...60R}, the [O~I] LVC is often structurally composite, consisting of a narrow component tracing a slow photoevaporative wind and a broad component centered near the stellar velocity that traces bound gas in Keplerian rotation. The exact morphology of the [O~I] profile is therefore tightly coupled to the stellar mass and incident radiation field. In lower-mass classical T Tauri stars, strong coronal X-ray and EUV emission can efficiently launch thermal photoevaporative winds, producing blueshifted narrow [O~I] components \citep{Ercolano2010MNRAS.406.1553E}. In contrast, intermediate-mass Herbig Ae stars like AB Aurigae lack deep convective envelopes and strong coronal emission, but their hotter photospheres emit substantially more FUV radiation. As a result, they are less effective at driving classical thermal photoevaporative outflows, and the [O~I] emission is more likely to be dominated by the broad, bound component \citep{Acke2005A&A...436..209A}.

Our high-resolution PARAS-2 observations of AB Aurigae closely follow this bound-disk picture. Using the adopted systemic velocity of $v_{\rm rad} = 15.1 \pm 0.1$ km s$^{-1}$, the [O~I] emission peaks at approximately the stellar rest velocity, with no significant blueshift. Given the nearly face-on inclination of $i \approx 22^\circ$, a vertical disk wind launched from the inner system would be expected to produce a measurable blueshift of several km s$^{-1}$, which is not observed. The absence of such a shift, combined with the symmetric line broadening, rules out a vertical outflow interpretation and instead indicates that the [O~I] line traces bound gas in Keplerian rotation. The measured FWHM of $\sim 35$ km s$^{-1}$  implies a projected rotational velocity of $\pm 17.5$ km s$^{-1}$, corresponding to a true Keplerian speed of $v_{\rm kep} \approx 46.7$ km s$^{-1}$. For a $2.4\,M_\odot$ central star, this places the [O~I]-emitting gas at a radial distance of order $1$ au.

This localization provides the missing kinematic link between the macro-scale late-stage infall mapped by ALMA (100 au scale) and the compact magnetospheric accretion traced by H$\alpha$ and He~I (sub-au scale). Recent ALMA observations have revealed large-scale streamers actively feeding fresh gas across the planet-carved cavity into the outer disk \citep{Speedie2025ApJ...981L..30S}. Our optical [O~I] detection shows that this material does not vanish in the cavity, but instead accumulates as a stable, rotating reservoir at the innermost edge of the thermally cleared dust-cavity region. From there, the gas is rapidly transported inward until it reaches the magnetospheric truncation radius, where it is funnelled onto the star. In this sense, the bound [O~I] reservoir acts as the direct physical intermediary between the ALMA-scale envelope infall and the extreme, clumpy magnetospheric accretion observed in the permitted lines.

Taken together, the [O~I], H$\alpha$, He~I, and Na~I diagnostics present a unified picture of AB Aurigae as a system in which large-scale external infall replenishes a compact inner reservoir, the reservoir feeds a severely crushed magnetosphere, and the magnetosphere in turn drives the observed accretion and wind variability. This makes AB Aurigae an especially valuable laboratory for understanding how disk gaps are crossed and how late-stage infall continues to sustain accretion in transition disks.

%%%%%%%conclusion 
\section{Conclusion} \label{sec:conclusion}

In this article, we present high-resolution ($R \approx 107{,}000$), multi-epoch optical spectroscopy of the Herbig Ae transition disk system AB Aurigae, using the PARAS-2 spectrograph mounted on the 2.5 m Mount Abu telescope. By resolving the complex temporal kinematics of the H$\alpha$, He~I $\lambda 5876$, [O~I] $\lambda 6300, 6363$ lines, alongside the Na~I~D doublet, we successfully map the physical structure of the innermost accretion environment ($r \lesssim 0.1$ au).

Our primary findings are as follows:

\begin{itemize}
\item Using our H$\alpha$ observations, we derive a highly active mass accretion rate of order $4 \times 10^{-7}\ M_\odot\ {\rm yr}^{-1}$, with values ranging from $(3.7 \pm 2.9)\times10^{-7}$ to $(4.0 \pm 3.2)\times10^{-7}\ M_\odot\ {\rm yr}^{-1}$. The infalling material is predominantly gaseous, as we find no spectroscopic signatures of large planetesimal or boulder infall in the Na~I~D profiles.

    \item The immense ram pressure from this accretion violently compresses the relatively weak stellar magnetic field ($B \lesssim 300$ G), driving the truncation radius inward to an extreme $R_{\rm mag} \approx 1.2 R_*$. This highly compact geometry coherently explains both the extreme rotational broadening in the H$\alpha$ wings ($v_{\rm kep} \approx 400$ km s$^{-1}$) and the severely restricted velocity dispersion of the He~I absorption (FWHM $\approx 160$ km s$^{-1}$). The magnetospheric accretion (MA) funnel flow is further confirmed by relatively stable velocities of $+32$ to $+34$ km s$^{-1}$ redshifted absorption dip in the H$\alpha$ profile.

    \item The system drives a highly variable, fast magnetocentrifugal wind ($-92$ to $-115$ km s$^{-1}$) that fluctuates on daily timescales, tightly coupled to the inner funnel flow. Conversely, a remarkably stable $-10.4$ to $-10.8$ km s$^{-1}$ low-velocity absorption dip traces an unperturbed photoevaporative or MHD disk wind launched further out in the extended disk.

    \item  Most importantly, we report the first detection of a persistent, single-peaked [O~I] emission profile centered at the stellar rest velocity, with a symmetric broadening of $\sim 35$ km s$^{-1}$. This emission does not trace an outflow or a redshifted infall stream; instead, it is consistent with a cool ($T \lesssim 3829$ K), gravitationally bound, stagnant Keplerian gas reservoir located at $\sim 1$ au.
\end{itemize}

\begin{figure*}[ht!]
    \centering
    \includegraphics[width=0.9\linewidth]{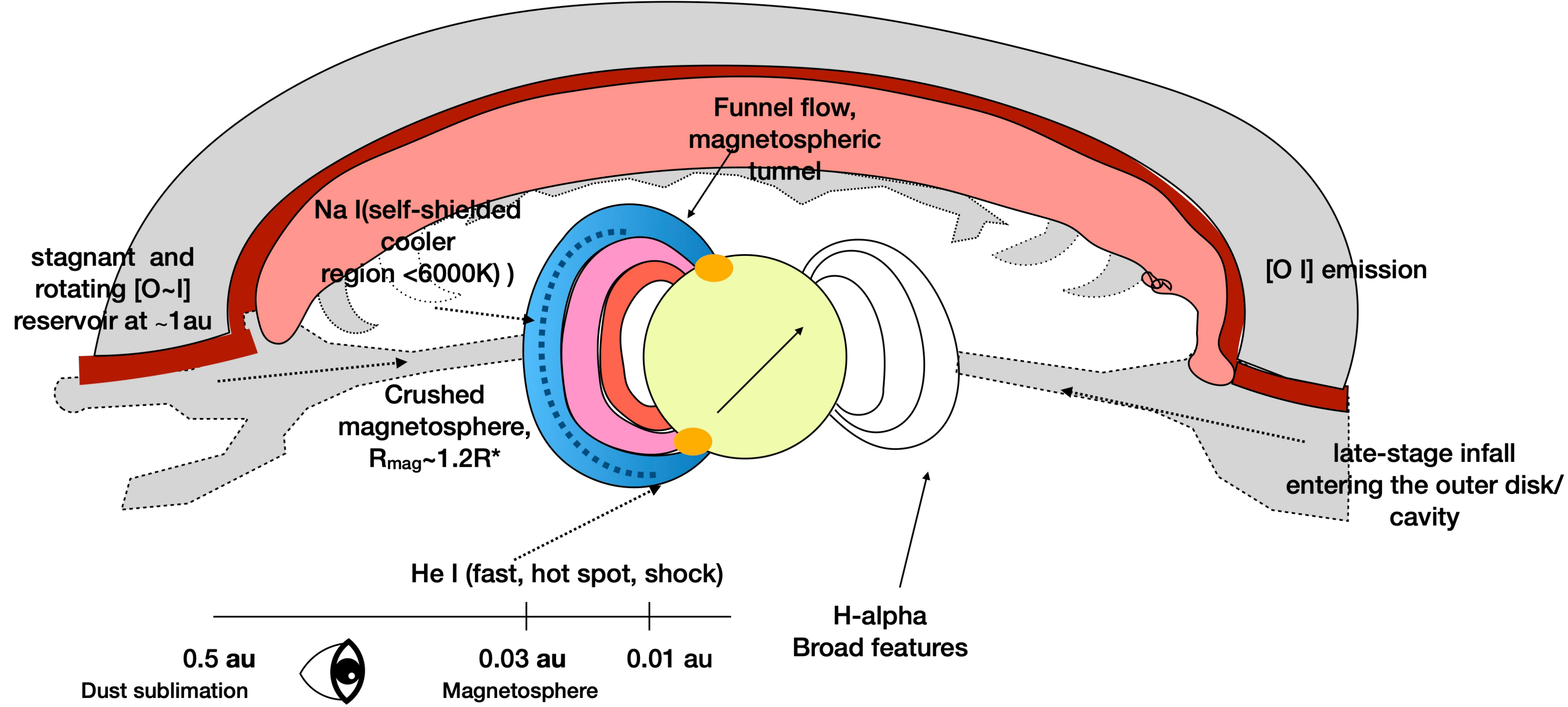}
    \caption{Schematic diagram of the multi-scale accretion environment in AB Aurigae. Gas from late-stage infall traverses the massive planet-carved outer dust ring (70--120 au) and accumulates as a bound [O~I] reservoir at $\sim 1$ au. This reservoir is traced by the symmetric [O~I] emission centered near the stellar rest velocity and serves as the missing intermediary between the macro-scale inflow and the compact inner magnetosphere. At the innermost boundary ($r \lesssim 0.03$ au), the immense ram pressure of the accretion flow crushes the stellar magnetic field to a compact truncation radius of $R_{\rm mag} \approx 1.2R_*$. The gas then couples to the magnetosphere, producing the broad He~I shock signature and the highly variable H$\alpha$ funnel-flow and wind components.}
    % \caption{\color{red} Schematic diagram of the multi-scale accretion environment in AB Aurigae. Gas from the late-stage infall traverses the massive, planet-carved outer dust ring ($70-120$ au) to replenish the inner system. The persistent [O I] +5 km s$^{-1}$1 emission comes from the far side and traces the cool ($< 3800$ K) gas inflow. At the innermost boundary ($r < 0.03$ au), the immense ram pressure of the accretion flow crushes the stellar magnetic field to a compact truncation radius of $R_{\rm mag} \approx 1.2 R_*$. The gas free-falls along these magnetospheric funnel flows, generating the broad, high-velocity H$\alpha$ emission and hot He~I shocks at the stellar surface. The massive columns of the funnel flow effectively absorb the stellar UV radiation, creating a self-shielded, cooler interior ($T < 6000$ K) where neutral sodium (Na~I) survives instant photoionization. All the related calculations and numbers mentioned here and presented in the main text can be found in Appendix. }
    \label{fig:schematic_diagram_AB_aurigae}
\end{figure*}

 Ultimately, these multi-tracer kinematics allow us to tie the entire physical picture together (see the schematic diagram in Fig.~\ref{fig:schematic_diagram_AB_aurigae}): the bound [O~I] reservoir acts as the missing intermediary between the macro-scale late-stage infall observed by ALMA and the heavily crushed $1.2 R_*$ magnetosphere. This overloaded boundary, in turn, powers both the highly variable free-falling funnel flow and the high-speed magnetocentrifugal exhaust. AB Aurigae thus serves as a powerful laboratory for showing how large-scale environmental infall is stored in a stagnant inner reservoir before directly driving extreme, micro-scale magnetospheric accretion and winds.

%% Please use the acknowledgment and contribution environments. This will 
%% be anonomyized when the "anonymous" style option is used. 
\begin{acknowledgments}
We are grateful to PRL-DOS (Department of Space, Government of India) and the Director, PRL, for their generous support. Their support has been instrumental in funding the PARAS-2 spectrograph. We express our gratitude to all the Mount Abu Observatory staff and the PARAS-2 instrument team for their invaluable support throughout the observations. D.S. acknowledges the support from the ANRF Ramanujan Fellowship for continuing research and PRL. We also acknowledge Dr. Vishal Joshi for sharing the telescope time.  J.C.-W. acknowledges funding by the European Union under the Horizon Europe Research \& Innovation Programme 101039452 (WANDA). Views and opinions expressed are, however, those of the author(s) only and do not necessarily reflect those of the European Union or the European Research Council. Neither the European Union nor the granting authority can be held responsible for them.
\end{acknowledgments}

%\begin{contribution}
%%This section gives authors the space to recognize author contributions. The text inside this environment is NOT counted towards the total word quanta. At a minimum, manuscripts are expected to include this text:

%All authors contributed equally to the Terra Mater collaboration.

%% But authors are expected to provide more specific details, e.g. 
%%
%%SC was responsible for writing and submitting the manuscript.
%%WWM came up with the initial research concept and edited the manuscript.
%%OTS obtained the funding and edited the manuscript.
%%EBF provided the formal analysis and validation. He also edited the manuscript.
%%GEH Supervised the undergraduates, wrote the software and administers the project github and Zenodo repositories.
%%
%% Authors can use the Contributor Role Taxonomy (CRediT) at
%% https://credit.niso.org
%% for ideas on how write a good statement tailored to their needs.

%\end{contribution}

%% To help institutions obtain information on the effectiveness of their 
%% telescopes the AAS Journals has created a group of keywords for telescope 
%% facilities.
%
%% Following the acknowledgments section, use the following syntax and the
%% \facility{} or \facilities{} macros to list the keywords of facilities used 
%% in the research for the paper.  Each keyword is check against the master 
%% list during copy editing.  Individual instruments can be provided in 
%% parentheses, after the keyword, but they are not verified.
\facilities{PARAS-2, Mount Abu 2.5 m telescope}

%% Similar to \facility{}, there is the optional \software command to allow 
%% authors a place to specify which programs were used during the creation of 
%% the manuscript. Authors should list each code and include either a
%% citation or url to the code inside ()s when available.
\begin{comment}

\software{astropy \citep{2013A&A...558A..33A,2018AJ....156..123A,2022ApJ...935..167A},  
          Cloudy \citep{2013RMxAA..49..137F}, 
          Source Extractor \citep{1996A&AS..117..393B}
          }
\end{comment}
%% Appendix material should be preceded with a single \appendix command.
%% There should be a \section command for each appendix. Mark appendix
%% subsections with the same markup you use in the main body of the paper.
%%
%% Each Appendix (indicated with \section) will be lettered A, B, C, etc.
%% The equation counter will reset when it encounters the \appendix
%% command and will number appendix equations (A1), (A2), etc. The
%% Figure and Table counter will not reset.

\appendix
\restartappendixnumbering

\section{Systemic Velocity and Kinematic Rest Frame} \label{app:Rv}

No sharp and stable photospheric absorption line is detected in our optical spectra that can be used to define an absolute rest frame for AB Aurigae. The apparent stellar radial velocity is known to vary due to intrinsic pulsations and other stellar activity, making the broad optical lines unreliable tracers of the true systemic velocity. This is consistent with the wide range of values reported in the literature, including the highly variable optical velocities discussed by \citet{Rodriguez2017RMxAC..49..166R}  and the older catalog value of $8.9$ km\,s$^{-1}$ adopted by \citet{Gontcharov2006AstL...32..759G}.

A more robust systemic velocity can be obtained from the stable outer molecular disk. Interferometric mapping of the Keplerian disk rotation has yielded $v_{\mathrm{LSR}} = 5.8 \pm 0.1$ km\,s$^{-1}$ from early IRAM observations \citep{Pietu2005A&A...443..945P}, while recent high-precision ALMA measurements of CS and SO gas constrain the disk kinematic center to $v_{\mathrm{LSR}} = 5.85 \pm 0.10$ km\,s$^{-1}$ \citep{Riviere2026A&A...707A.348R}. To place our optical spectra in the true rest frame of the star--disk system, we converted this LSR velocity to the barycentric frame using
\[
v_{\mathrm{rad}} = v_{\mathrm{LSR}} - V_{\mathrm{proj}},
\]
where the projected solar motion is defined by spherical trigonometry:
\[
V_{\mathrm{proj}} = V_{\odot}
\left[
\sin\delta \sin\delta_{\mathrm{apex}}
+
\cos\delta \cos\delta_{\mathrm{apex}}
\cos(\alpha - \alpha_{\mathrm{apex}})
\right].
\]

Assuming the standard radio kinematic solar motion of $V_{\odot} = 20.0$ km\,s$^{-1}$ directed toward the solar apex at $\alpha_{\mathrm{apex}}=18^{\mathrm h}$ and $\delta_{\mathrm{apex}}=+30^\circ$, and adopting the J2000 coordinates of AB Aurigae $(\alpha = 04^{\mathrm h}55^{\mathrm m}45.8^{\mathrm s},\ \delta = +30^\circ33'04'')$, we obtain an exact geometrical projection of $V_{\mathrm{proj}} = -9.25$ km\,s$^{-1}$. 

Because the standard kinematic LSR is a defined reference frame and the stellar coordinates are known to extreme precision, the uncertainty in this geometric transformation is negligible. Therefore, the uncertainty in the final barycentric velocity is entirely dominated by the millimeter observational error. Applying this correction to the ALMA systemic velocity yields:
\[
v_{\mathrm{rad}} = 15.1 \pm 0.1 \; \mathrm{km\,s^{-1}}.
\]
We adopt $15.1 \pm 0.1$ km\,s$^{-1}$ as the definitive systemic barycentric radial velocity of the AB Aurigae system throughout this work.

As an independent internal check on the adopted systemic velocity, we note that the low-velocity H$\alpha$ Central Dip (Section~3.3) is observed at $-10.4$ to $-10.8$ km s$^{-1}$ in the stellar rest frame. This velocity is consistent with the isothermal sound speed expected for gas at the photospheric temperature of AB Aurigae ($T_{\rm eff} \approx 9500$ K; $c_{\rm s} \approx 11$ km s$^{-1}$), supporting a thermally driven photoevaporative wind origin. The close agreement between the observed Central Dip velocity and the expected sound speed therefore provides an independent consistency on the adopted systemic velocity.

\section{ Detection and Estimation  of [O I] line luminosity and excitation temperature limits}
\label{sec:appendix_oi_temp}

 %%%%%%%%%%%%%%%%%%%%%%%%%%%%%%%%%

\subsection{[O I] detection}
\begin{figure*}
    \centering
    \includegraphics[width=0.6\linewidth]{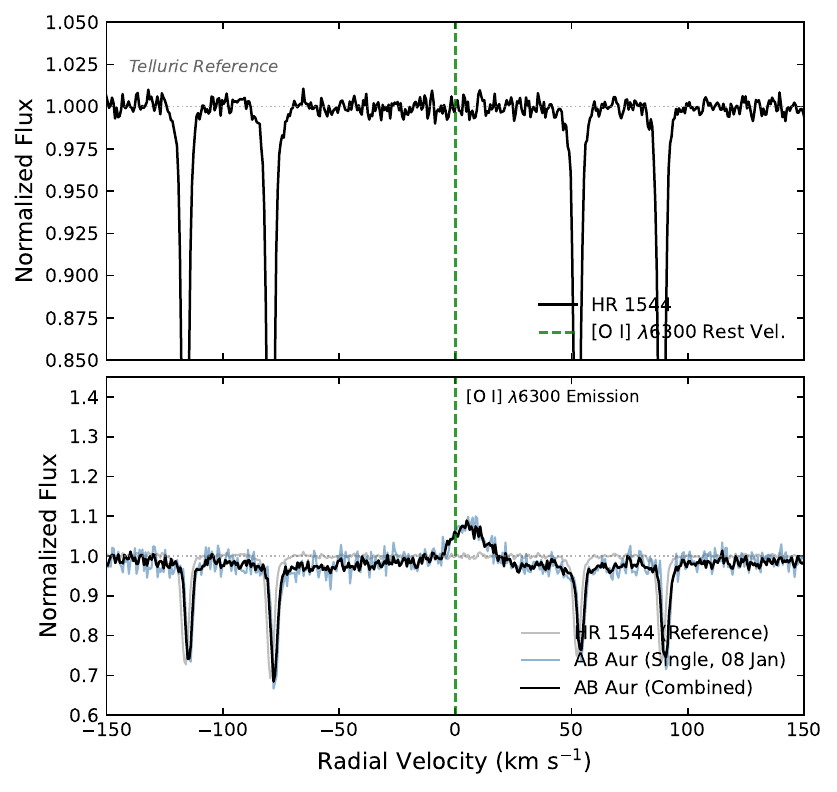}
    \caption{Comparison spectra of standard star and AB Aurigae, this clearly shows the presence of forbidden oxygen line emission towards AB Aurigae. }
    \label{fig:combined_spectra}
\end{figure*}

Initial inspection of the barycentric-corrected PARAS-2 spectra revealed potential forbidden [O~I] $\lambda$6300, 6363\,\AA\ emission toward AB Aurigae. Because this wavelength regime is highly susceptible to telluric absorption and atmospheric [O~I] airglow, a rigorous telluric correction was required to confirm the circumstellar origin of the emission. Following blaze correction and continuum normalization, we used the rapidly rotating A1V standard star HR~1544 to map the Earth's atmospheric transmission. Although the standard star integration time (900\,s) was half that of the target (1800\,s), we compensated for the resulting $\approx$1.4$\times$ sensitivity deficit by co-adding the normalized standard spectra across all three epochs (January 8, 9, and 10). The resulting combined spectrum, shown in the top panel of Figure~\ref{fig:combined_spectra}, clearly traces many narrow telluric absorption features while no spectral feature {\bf near} 6300\,\AA was observed.

The telluric-corrected AB Aurigae spectra were obtained by direct division of the target by the standard star spectrum. This approach cleanly removed all telluric contamination without significant residuals. The bottom panel of Figure~\ref{fig:combined_spectra} displays the corrected spectra alongside the standard star reference. To verify that co-adding epochs did not artificially smear the [O~I] profile, we compared a single-epoch 30-minute exposure with the co-added spectrum; both show  morphological agreement, confirming the robustness of the [O I] detection.

%%%%%%%%%%%%%%%%%%%%%%%%%%%%%%%%%%%%%%

\subsection{Absolute Line Luminosity}
To constrain the physical conditions of the [O~I] emitting region in AB Aurigae, we calculated the absolute line luminosity of the [O~I] $\lambda$6300 transition and placed an upper limit on the gas excitation temperature using the non-detection of the higher-excitation [O~I] $\lambda$5577 transition. The calculations were performed on the telluric-corrected, continuum-normalized spectra, which were shifted to the stellar rest frame.

The equivalent width (EW) of the [O~I] $\lambda$6300 emission line was estimated using numerical integration across the line profile. For the January 8 epoch, this gave $\mathrm{EW}_{6300} \approx 0.1780 \pm 0.006\,\AA$, and this is representative of the other epochs as well.

 To convert this normalized EW into an absolute line flux ($F_{\mathrm{line}}$), we bypassed broad-band photometric approximations and estimated the intrinsic continuum flux directly using a stellar photosphere model. We adopted a blackbody profile for AB Aurigae with $T_{\rm eff} = 9500$\,K and $R_\star = 2.5\,R_\odot$, scaled by the Gaia EDR3 geometric distance of $d = 162.9 \pm 1.5$\,pc. Evaluating this model at exactly $6300.3$\,\AA\ yields an unreddened continuum flux of $F_{\mathrm{cont},6300} \approx 4.48 \times 10^{-12}$ erg s$^{-1}$ cm$^{-2}$ \AA$^{-1}$.

Multiplying the measured EW by this continuum flux gives an intrinsic line flux of $F_{\mathrm{line},6300} \approx 7.98 \times 10^{-13}$ erg s$^{-1}$ cm$^{-2}$. Using the Gaia EDR3 distance of $d = 162.9$ pc, we derive an absolute line luminosity of
\begin{equation}
    L_{6300} = 4 \pi d^2 F_{\mathrm{line},6300} \approx 6.62 \times 10^{-4} \, L_\odot.
\end{equation}

\subsection{Excitation Temperature Upper Limit}
We have not detected the [O~I] $\lambda$5577 line, which comes from a higher excitation state ($^1S$) than the $\lambda$6300 line ($^1D$). The flux ratio of these two lines is highly sensitive to the gas temperature and electron density, which acts as a classic diagnostic of the emitting region's thermal state.

% We calculated a $3\sigma$ upper limit for the $\lambda$5577 EW assuming the same kinematic line width (FWHM $\approx 32$ km s$^{-1}$) as the $\lambda$6300 line. Using the local continuum noise level ($\sigma \approx \mathrm{0.005}$), we established an upper limit of $\mathrm{EW}_{5577} < \mathrm{0.028}$\,\AA.

 The continuum ratio from the blackbody model is $F_{\mathrm{cont},6300}/F_{\mathrm{cont},5577} \approx 0.762$, and the non-detection of [O~I] $\lambda5577$ yields a $3\sigma$ upper limit of $\mathrm{EW}_{5577} < 0.0283$\,\AA. This implies a lower limit on the line luminosity ratio of $L_{6300}/L_{5577} > 4.8$.

Assuming Local Thermodynamic Equilibrium (LTE) and high electron densities ($n_e > 10^6$ cm$^{-3}$, typical of inner disk environments) where collisional de-excitation dominates, the temperature can be constrained by the relation \citep[e.g.,][]{Gorti2011ApJ...735...90G}:
\begin{equation}
    \frac{L_{6300}}{L_{5577}} = 5.4 \times 10^{-3} \exp\left(\frac{26000}{T}\right)
\end{equation}
Inserting our empirical lower limit for the line ratio yields a maximum gas excitation temperature of
\begin{equation}
    T_{\mathrm{max}} = \frac{26000}{\ln \left( \frac{L_{6300}/L_{5577}}{5.4 \times 10^{-3}} \right)} \approx 3829\,\text{K}
\end{equation}
This relatively low temperature upper limit confirms that the [O~I] emission does not originate in an ultra-hot, highly ionized accretion shock at the stellar surface. Instead, it is entirely consistent with a cooler, gravitationally bound, stagnant gas reservoir pooling within the inner cavity prior to magnetospheric accretion.

%
%%%%%%%Figure moved to appendix section 

%%%%%%%%%%%================================
\section{Mass accretion rate calculation from H$\alpha$ }
\label{sec:appendix_macc}

To estimate the mass accretion rate (\(\dot{M}_{\rm acc}\)) from the H\(\alpha\) emission line, we followed the luminosity conversion method calibrated for Herbig Ae/Be stars by \citet{Fairlamb2015MNRAS.453..976F} (see Table 2 there) .

First, the observed H\(\alpha\) equivalent width was converted to a line luminosity (\(L_{{\rm H}\alpha}\)), adopting a Gaia distance of \(d = 163\)~pc and a visual extinction of \(A_V = 0.5\)~mag (\(A_{{\rm H}\alpha} \approx 0.4\)~mag). The total accretion luminosity (\(L_{\rm acc}\)) was then derived using the empirical relationship:
\begin{equation}
    \log(L_{\rm acc}/L_\odot) = 1.00 \log(L_{{\rm H}\alpha}/L_\odot) + 2.09
    \label{eq:L_acc}
\end{equation}

The mass accretion rate was subsequently calculated using the standard relation:
\begin{equation}
    \dot{M}_{\rm acc} = \frac{L_{\rm acc} R_*}{G M_* \left(1 - \frac{R_*}{R_{\rm in}}\right)}
    \label{eq:M_acc}
\end{equation}
where \(G\) is the gravitational constant, \(M_* = 2.4\,M_\odot\), and \(R_* = 2.5\,R_\odot\) .

The choice of the inner disk truncation radius (\(R_{\rm in}\)) is critical for Herbig Ae stars. While standard population surveys often adopt \(R_{\rm in} = 5.0\,R_*\) for consistency with low-mass T Tauri stars \citep[e.g.,][]{Fairlamb2015MNRAS.453..976F}, kinematic studies of Herbig Ae stars demonstrate they possess highly compact magnetospheres due to their weaker magnetic fields \citep{Mendigutia2011A&A...535A..99M, Cauley2014ApJ...797..112C}. Indeed, our own kinematic estimates (Appendix C) yield a truncation radius of \(R_{\rm mag} \approx 1.2 R_*\) . However, utilizing \(R_{\rm in} \approx 1.2 R_*\) in Equation \ref{eq:M_acc} diverges significantly from the foundational assumptions used to calibrate Equation \ref{eq:L_acc}. Therefore, to remain strictly consistent with the \citet{Fairlamb2017MNRAS.464.4721F} empirical framework, we adopt the standard \(R_{\rm in} = 5.0\,R_*\) for the final \(\dot{M}_{\rm acc}\) calculation, noting that this represents a lower limit; a more compact physical magnetosphere would yield an even higher accretion rate.

 Using the empirical calibrations of \citet{Fairlamb2015MNRAS.453..976F}, we derive a mass accretion rate of order $\dot{M}_{\rm acc} \approx 4 \times 10^{-7}\,M_\odot\,{\rm yr}^{-1}$ from our PARAS-2 H$\alpha$ profiles. This value is modestly higher than the historical baseline estimates for AB Aurigae, which are of order $\sim 1.3 \times 10^{-7}\,M_\odot\,{\rm yr}^{-1}$ \citep{Garcia2006A&A...459..837G, Salyk2013ApJ...769...21S}. Given the known variability of accretion in young stellar objects, such a difference is not unexpected and is consistent with the strongly time-variable accretion state of AB Aurigae during our observing window, as evidenced by the night-to-night changes in the H$\alpha$ and He~I line profiles.

The mass accretion rates derived from our 2026 PARAS-2 observations are consistent with the broader picture of episodic, clumpy accretion in Herbig Ae systems. Accretion rates in young stellar objects are known to vary by $0.3$--$0.5$ dex over time due to changes in the accretion geometry and variable mass loading of the inner flow \citep[e.g.,][]{Bouret1998A&A...340..163B, Costigan2012MNRAS.427.1344C, Mendigutia2011A&A...535A..99M}. In this context, the elevated accretion rate inferred here supports our conclusion that AB Aurigae is currently experiencing an actively fed accretion state, plausibly linked to late-stage infall traversing the planet-carved cavity.

\section{ magnetospheric truncation and funnel flow kinematics}
\label{sec:appendix_kinematics}

To physically interpret the extreme kinematics observed in the H\(\alpha\) and He~I profiles of AB Aurigae, we modeled the magnetospheric accretion geometry using standard analytical formulations \citep[e.g.,][]{Bouvier2007prpl.conf..479B, Hartmann2016ARA&A..54..135H}. 

Assuming the stellar magnetic field is primarily dipolar, the magnetospheric truncation radius (\(R_{\rm mag}\))---the point where the magnetic pressure balances the ram pressure of the infalling disk material---is given by:
\begin{equation}
    R_{\rm mag} = \left( \frac{\mu^2}{\dot{M}_{\rm acc} \sqrt{2GM_*}} \right)^{2/7}
    \label{eq:R_mag}
\end{equation}
where \(\mu = B R_*^3\) is the stellar magnetic dipole moment. For AB Aurigae, we adopted a typical upper-limit surface magnetic field for Herbig Ae stars of \(B \approx 300\)~G \citep[e.g.,][]{Alecian2013MNRAS.429.1001A, Wade2007MNRAS.376.1145W}, alongside our derived mass accretion rate (\(\dot{M}_{\rm acc} \approx 4.0 \times 10^{-7}\,M_\odot\,{\rm yr}^{-1}\)), \(M_* = 2.4\,M_\odot\), and \(R_* = 2.5\,R_\odot\). This yields a highly compact truncation radius of \(R_{\rm mag} \approx 1.2 R_*\).

Gas lifted from the disk at \(R_{\rm mag}\) rotates at the local Keplerian velocity (\(v_{\rm kep}\)) before accelerating along the magnetic field lines. The Keplerian velocity at the inner disk edge is:
\begin{equation}
    v_{\rm kep} = \sqrt{\frac{GM_*}{R_{\rm mag}}}
    \label{eq:v_kep}
\end{equation}
which yields \(v_{\rm kep} \approx 400\)~km~s\(^{-1}\). 

As the gas follows the magnetic funnel flow to the stellar surface, it reaches a maximum free-fall velocity (\(v_{\rm ff}\)) dictated by the change in gravitational potential:
\begin{equation}
    v_{\rm ff} = \sqrt{2GM_* \left( \frac{1}{R_*} - \frac{1}{R_{\rm mag}} \right)}
    \label{eq:v_ff}
\end{equation}
Due to the compact nature of the magnetosphere (\(R_{\rm mag} \approx 1.2 R_*\)), the gas only accelerates over a short distance, resulting in a maximum line-of-sight free-fall velocity of \(v_{\rm ff} \approx 230\)~km~s\(^{-1}\). 

These derived kinematic parameters are consistent with our observational data. 
The He~I absorption profile traces gas falling directly toward the stellar 
photosphere. In a standard dipole geometry ($r = R_{\rm mag} \sin^2 \theta$), 
a highly compact magnetosphere with $R_{\rm mag} \approx 1.2 R_*$ forces the 
accreting gas to follow shallow magnetic field lines. The flow then impacts 
the stellar surface at low magnetic latitudes ($\theta_* \approx 66^\circ$). 
Because the AB Aurigae system is viewed nearly face-on ($i \approx 22^\circ$; \citealt{Tang2017ApJ...840...32T}), 
this funnel flow is projected largely transverse to our line of sight. 
Consequently, the theoretical three-dimensional terminal free-fall velocity 
($v_{\rm ff} \approx 230$ km s$^{-1}$) is reduced by geometric projection to 
an observable line-of-sight redshift of $v_{\rm obs} \approx 130$ km s$^{-1}$. 
This is consistent with the measured terminal edge of our He~I profile. 
Furthermore, the velocity gradient along this trajectory, convolved with local 
post-shock thermal and turbulent broadening, further broadens the line width 
(FWHM $\approx 160$ km s$^{-1}$).

%These derived kinematic parameters perfectly reproduce our observational data. 
%he He~I absorption profile traces gas falling directly against the stellar 
%photosphere. In a standard dipole geometry ($r = R_{\rm mag} \sin^2 \theta$), 
%{\color {red} a highly compact magnetosphere of $R_{\rm mag} \approx 1.2 R_*$ forces the 
%accreting gas to follow shallow magnetic field lines. Rather than falling into 
%the magnetic poles, the gas impacts the stellar surface at low magnetic latitudes 
%($\theta_* \approx 66^\circ$). Because the AB Aurigae system is viewed nearly 
%face-on ($i \approx 22^\circ$; \citealt{Tang2017ApJ...840...32T}), this 
%near-equatorial infall is directed largely transverse to our line of sight. 
%Consequently, the theoretical 3D terminal free-fall velocity ($v_{\rm ff} \approx 230$ 
%km s$^{-1}$) is subject to a strong geometric projection effect, naturally reducing the maximum observable line-of-sight redshift to $v_{\rm obs} \approx 130$ km s$^{-1}$. This is in good quantitative agreement with the measured terminal edge of our He~I profile. Furthermore, the velocity gradient along this trajectory, convolved with local post-shock thermal and turbulent broadening, further broaden the line width (FWHM $\approx 160$ km s$^{-1}$).}

%%%%%%%%%%%discussion on sodium 
\section{Sodium (Na) Magnetospheric Infall Signatures}
\label{sec:appendix_na}

A critical question may arise regarding the physical origin of these Na~I absorption features. Given the intense radiation field and high effective temperature ($T_{\rm eff} \approx 9500$ K) of the central A-type star, unshielded neutral sodium cannot naturally survive in the inner circumstellar environment. To address this, we systematically evaluate two distinct physical scenarios to determine the most plausible origin for the observed sodium gas. \\

{\bf Scenario 1: Boulder/planetesimal infall:}
We can estimate the minimum planetesimal size required to produce a Na~I absorption feature with a normalized depth of 0.25 and a width of 25 km s$^{-1}$, consistent with our observations. In case where lines are optically saturated, the unshielded Na~I column density is approximately $N_{\text{Na I}} \approx 10^{12} \text{ cm}^{-2}$ \citep{deWinter1999A&A...343..137D}. Because adding more sodium to a saturated line does not significantly increase its absorption depth, this calculation strictly provides a lower limit for the required physical size of the body. By adopting a standard planetesimal bulk density of $2 \text{ g cm}^{-3}$ and assuming a cosmic sodium abundance, generating a gas cloud capable of covering 25\% of the stellar photosphere requires a pure sodium mass of $\approx 9 \times 10^{11} \text{ g}$. This corresponds to a total sublimated rock mass of $\approx 3 \times 10^{16} \text{ g}$, yielding a physical radius of 1.5 km (or a boulder diameter of $\approx 3 \text{ km}$). However, this calculation inherently assumes that all sodium released from the evaporating boulder remains neutral at a distance of 0.1 au from the hot A-type star. This is highly unlikely, as $>99\%$ of the unshielded sodium would be instantly photoionized into Na~II by the intense stellar radiation field, making it invisible in these observations \citep{Sorelli1996A&A...309..155S}. Furthermore, theoretical evaporation models demonstrate that to survive this intense radiation field and penetrate down to 0.1 au without disintegrating entirely, a solid body must be at least 100 km in diameter \citep{Beust2001A&A...366..945B, Chakraborty2004ApJ...606L..69C}. 

Observationally, instantaneous sodium desorption from an evaporating solid body would produce a highly localized, narrow absorption profile. For gas at $T \sim 6000$ K, thermal broadening is only $\Delta v_{\rm thermal} \approx 2 \text{ km s}^{-1}$ \citep{Muzerolle2004ApJ...617..406M}. Because small pebbles sublimate immediately upon crossing the dust destruction radius, their released gas cannot survive photoionization long enough to traverse the hot inner radiation environment as neutral sodium. Therefore, it would not  generate the broad ($\sim 25 \text{ km s}^{-1}$) macroscopic velocity gradients seen in our spectra. Furthermore, global disk constraints strongly disfavor a continuous stream of massive boulders. The presence of giant planets within the cavity creates pressure bumps that may trap large solids at the outer disk edge ($\approx 70-120$ au), physically halting their inward drift into the inner system. Even if large boulders could efficiently bypass this trap, sustaining our derived mass accretion rate entirely through solid bodies would deplete the estimated total mass of the AB Aurigae disk ($M_{\rm disk} \approx 0.01\, M_\odot$; \citealt{DeWarf2003ApJ...590..357D, Andrews2005ApJ...631.1134A}) in only $\sim 10^5$ years. Given that AB Aurigae is an evolved $1-4$ Myr old system exhibiting continuous, long-term accretion \citep{DeWarf2003ApJ...590..357D}, such a rapid depletion timescale is unsustainable. Therefore, unless we have fortuitously observed an extremely rare, transient infall of a 100-km planetesimal \citep[e.g.,][]{Chakraborty2004ApJ...606L..69C}, continuous pebble or boulder accretion cannot physically explain the observed Na~I~D signature. \\

{\bf Scenario 2: Na~I~D absorption from magnetospheric accretion:} The observed FWHM of the Na~I absorption troughs in our spectra ($\approx 20-25$ km s$^{-1}$) significantly exceeds the expected thermal broadening for gas at $T \sim 6000$ K ($\Delta v_{\rm thermal} \approx 2$ km s$^{-1}$; \citealt{Muzerolle2004ApJ...617..406M}). This implies that the line width is dominated by bulk kinematic broadening. Such a velocity dispersion is entirely consistent with the macroscopic velocity gradients expected across the radial extent of a magnetospheric funnel flow accelerating towards the star. In order for neutral sodium to survive the intense UV radiation field at these high-velocity, close-in distances ($\lesssim 0.1$\,au), it must be protected by the massive column density of the funnel flow itself. The core of the magnetospheric accretion columns reaches extreme densities ($n_H \approx 10^{11} - 10^{12} \text{ cm}^{-3}$; \citealt{Ingleby2013ApJ...767..112I, Hartmann2016ARA&A..54..135H}). The outer layers of these dense streams effectively absorb the destructive stellar UV photons, providing a self-shielded, cooler ($T \lesssim 6000$ K) interior where Na~I can remain neutral and be observed in absorption against the stellar continuum.

Furthermore, the coherent correlation of the Na~I absorption depth with our derived mass accretion rate, alongside its kinematic alignment with the He~I centroid at $\approx +30$ km s$^{-1}$, are established hallmarks of magnetospheric gas \citep{Muzerolle2004ApJ...617..406M, Rebollido2020A&A...639A..11R}. This behavior stands in stark contrast to spectroscopic signatures of planetesimal infall---such as those observed in the Herbig Be star LkH$\alpha$ 234---which are characterized by highly localized, narrow ($\lesssim 5$ km s$^{-1}$), episodic, and kinematically transient absorption events that are completely uncorrelated with the bulk H$\alpha$ and He~I accretion rates\citep{Chakraborty2004ApJ...606L..69C}. %\citep{2000ASPC..219..376G, 2004ApJ...606L..73C}. 

Therefore, the broad,  stable, and accretion-rate-correlated Na~I absorption observed in AB Aurigae could be attributed to a continuously replenished self-shielded magnetospheric gas reservoir. Rather than arising from the episodic sublimation of scattered planetesimals, this sodium feature provides a direct, stable kinematic map of the macroscopic funnel flow.

%% For this sample we use BibTeX plus aasjournalv7.bst to generate the
%% the bibliography. The sample7.bib file was populated from ADS. To
%% get the citations to show in the compiled file do the following:
%%
%% pdflatex sample7.tex
%% bibtext sample7
%% pdflatex sample7.tex
%% pdflatex sample7.tex

\bibliography{sample701}{}
\bibliographystyle{aasjournalv7}

%% This command is needed to show the entire author+affiliation list when
%% the collaboration and author truncation commands are used.  It has to
%% go at the end of the manuscript.
%\allauthors

%% Include this line if you are using the \added, \replaced, \deleted
%% commands to see a summary list of all changes at the end of the article.
%\listofchanges

\end{document}